\documentclass[AMA,STIX1COL]{WileyNJD-v2}
\usepackage{moreverb}
\usepackage{makecell}
\usepackage{multicol}
\usepackage{multirow}
\usepackage{bbm}
\usepackage[table]{xcolor}
\usepackage{float}
\usepackage{placeins}

\newcommand\BibTeX{{\rmfamily B\kern-.05em \textsc{i\kern-.025em b}\kern-.08em
T\kern-.1667em\lower.7ex\hbox{E}\kern-.125emX}}

\articletype{Research Article}%

\received{<day> <Month>, <year>}
\revised{<day> <Month>, <year>}
\accepted{<day> <Month>, <year>}

\begin{document}

\title{Accessibility and Serviceability Assessment 
to Inform Offshore Wind Energy Development and Operations off the U.S. East Coast}
\author[1]{Cory Petersen}
\author[2]{Feng Ye}
\author[1]{Jiaxiang Ji}
\author[3]{Josh Kohut}
\author[1,4]{Ahmed Aziz Ezzat$^*$}
\author[5]{David Saginaw}
\author[5]{Avril Montanti}
\author[5]{Jack Cammarota}

\authormark{Petersen \textsc{et al}}

\address[1]{\orgdiv{Department of Industrial \& Systems Engineering}, \orgname{Rutgers, The State University of New Jersey}, \orgaddress{\state{NJ}, \country{United States}}}

\address[2]{\orgdiv{Department of Industrial Engineering}, \orgname{Clemson University}, \orgaddress{\state{SC}, \country{United States}}}

\address[3]{\orgdiv{Department of Marine \& Coastal Sciences}, \orgname{Rutgers, The State University of New Jersey}, \orgaddress{\state{NJ}, \country{United States}}}

\address[4]{\orgdiv{University College}, \orgname{Korea University}, \orgaddress{\state{Seoul}, \country{Republic of Korea}}}

\address[5]{\orgdiv{Blue Ocean Transfers}, \orgname{McQuilling Renewables}, \orgaddress{\state{NY}, \country{United States}}}


\corres{$^*$Ahmed Aziz Ezzat, 96 Frelinghuysen Rd, Piscataway, NJ, 08854, USA 
\email{aziz.ezzat@rutgers.edu}}


\abstract[Abstract]{
The economic success of offshore wind energy projects relies on accurate projections of the construction, and operations and maintenance (O\&M) costs. These projections must consider the logistical complexities introduced by adverse met-ocean conditions that can prohibit access to the offshore assets for sustained periods of time. In response, the goal of this study is two-fold: (1) to provide high-resolution estimates of the accessibility of key offshore wind energy areas in the United States (U.S.) East Coast\textemdash a region with significant offshore wind energy potential; 
and (2) to introduce a new operational metric, called \textit{serviceability}, as motivated by the need to assess the accessibility of an offshore asset along a vessel travel path, rather than at a specific site, as commonly carried out in the literature. We hypothesize that serviceability is more relevant to offshore operations than accessibility, since it more realistically reflects the success and safety of a vessel operation along its journey from port to site and back. 
Our analysis reveals high temporal and spatial variations in accessibility and serviceability, even for proximate offshore locations. We also find that solely relying on numerical met-ocean data can introduce considerable bias in estimating accessibility and serviceability, raising the need for a statistical treatment that combines both numerical and observational data sources, such as the one proposed herein. 
Collectively, our analysis sheds light on the value of high-resolution met-ocean information and models in supporting offshore operations, including but not limited to future offshore wind energy developments.} 


\maketitle


\section{Introduction}\label{sec1}
Driven by an increasing global demand for energy, coastal nations are turning to offshore energy resources in an effort to unlock access to ocean-based abundant energy resources while advancing the economic development and sustainable growth of their coastal economies. Among these ocean-based resources, 
offshore wind energy is projected to grow rapidly into a major source of global renewable energy generation, owing to its numerous environmental and economic benefits \cite{musial2023offshore}. 
The United States (U.S.) in particular has significant offshore wind potential, 
distributed across a number of geographical regions off of its coastline \cite{musial2010large}. Of these, the U.S. East Coast is expected to be the first and largest contributor to future offshore wind developments due to 
its relatively shallow waters, steady winds, and proximity to major coastal population centers\cite{NOWRDC2024}. These characteristics make the East Coast the most feasible and economically promising region in the United States for offshore wind development in the near term \cite{gallaher2023breaking,njdep}, and hence, is the geographical focus of this work.  

To enable the economic success of future offshore wind projects and de-risk their development and operations, there is a need for accurate, region-specific projections of the construction, operations and maintenance (O\&M) requirements and costs
\cite{musial2022offshore, sheng2023wind,maples2013installation}. For example, O\&M activities alone contribute $\sim$\hspace{-0.35mm}$30$-$40$\% of an offshore wind project's total life cycle costs\cite{stehly20202018}. Thus, efforts to precisely estimate O\&M expenditures, let-alone to reduce them, can have a significant impact on the future economic prospect of an offshore wind project. However, obtaining accurate offshore wind O\&M projections is not trivial, in part due to the logistical complexities introduced by a dynamic met-ocean environment which can significantly hinder the capability of wind farm operators to effectively and safely schedule, dispatch, and undertake critical construction and O\&M operations at sea\cite{dalgic2015advanced, dinwoodie2015reference}. 

Unlike land-based wind farms, operational activities in an offshore environment are largely restricted by adverse met-ocean conditions. If these conditions exceed their perceived safety limits, the ability to transport personnel, material, and equipment to the offshore assets is prohibited. As a result, time-sensitive maintenance actions may be delayed until milder conditions arise, which can have a significant impact on the availability of offshore assets, especially when considering sensitive or costly repairs\cite{elsahhar2025analysis, carroll2016failure}. 
Recent studies suggest that properly modeling met-ocean conditions in a wind farm can help reduce the total O\&M costs of an offshore wind farm by as much as $7$\%\cite{petros2021, petros2023}. This is because accurate information about met-ocean conditions can enable operators to proactively schedule maintenance tasks at times of higher likelihood of success,
thereby reducing unnecessary delays due to asset shutdowns and maintenance interruptions, ultimately lowering the overall O\&M costs\cite{catterson2016economic, petros2022}. 

Recognizing its impact on offshore construction and O\&M planning, there have been prior efforts to examine the accessibility of offshore locations, with the overwhelming majority of these efforts focused on offshore wind energy projects in the North Sea\cite{martini2017accessibility, van2003analysis, catterson2016economic, silva2013impact}. 
Unique to our study are the following aspects:
\begin{itemize}
\item 
Up to the authors' knowledge, this study presents the {first} attempt to provide high-resolution accessibility estimates for the potential offshore wind energy areas off the U.S. East Coast. 
We envision our findings to provide timely insights to the developers and operators of future and current large-scale offshore wind farms in this region, guiding their construction and O\&M projections and supporting their logistical planning decisions. This is especially relevant considering that vessel shortage has been cited as a major bottleneck for offshore wind energy development in the United States and elsewhere\cite{vesselshortage} and hence, efficient utilization of vessel resources which considers high-fidelity and precise logistical information about accessibility will be key to enable the economic success of the offshore wind industry \cite{shields2022demand}.  
    
\item The majority of publicly available accessibility assessment studies primarily rely on numerical model data to derive accessibility estimates\cite{van2003dowec,silva2013impact,martini2017accessibility,rowell2022does}. This is because long streaks of spatio-temporal observational data, which are needed for accurate accessibility estimation, are typically unavailable, especially for new offshore wind developments. Observing the inherent biases of numerical models, we propose a statistical regression framework to combine observations that are scarce (temporally) and sparse (spatially), with numerical model data, in order to derive accurate, high-resolution, estimates of accessibility that cannot be derived by solely relying on observational or numerical model data alone. 

\item We propose a new logistical metric, which we call \textit{serviceability}, aiming to assess the accessibility of an offshore site along a vessel's travel path, and considering the vessel's operational profile. This is unlike the commonly adopted definition of accessibility which only considers the met-ocean conditions at an offshore site, and can lead to overly conservative logistical decisions, thereby forfeiting work opportunities due to imprecise met-ocean assessments. We hypothesize that serviceability is more relevant to construction and O\&M planning than accessibility, since it more realistically reflects the success of the offshore activities 
throughout the offshore mission, from port to site.  

\item We carry out season-to-season, and site-to-site comparisons of accessibility and serviceability for key offshore energy regions off the U.S. East Coast. This analysis reveals important spatial and temporal variability in accessibility and serviceability, which can provide crucial insights to developers of future offshore energy projects in de-risking their economic bids (pre-leasing), as well as determining their logistical requirements during construction, operations, and maintenance. 

\end{itemize}

The paper is organized as follows. Section 2 describes the data used in this work, and its relevance to the U.S. offshore wind energy sector. Section 3 introduces the proposed accessibility estimation framework combining observational and numerical model data, and introduces serviceability as a new logistical metric to assess accessibility along a vessel route. Section 4 presents the results of our accessibility and serviceability assessment off the U.S. East Coast, and discusses relevant analyses and findings to the offshore wind areas therein. Finally, Section 5 concludes the paper, outlines future research directions, and discusses the applicability of the methods and analyses conducted herein to broader industries and applications of the Blue Economy.

\section{Data Description and Exploratory Analysis} 
Several factors can impact the accessibility of an offshore asset. These can be broadly categorized into: 
\begin{itemize}
    \item \textit{Environmental factors}, which refer to the met-ocean conditions (i.e., sea and weather states) that can restrict the safe access of vessels, equipment, material, and crew, to an offshore asset. 
    \item \textit{Operational factors}, which refer to the logistical resources available (e.g., mode of transport, access system), as well as operational parameters related to the offshore missions to be carried out (e.g., mission duration, complexity, vessel's expected travel path and operational profile). 
\end{itemize}
In what follows, we proceed to describe each of these sets of factors. 


\subsection{Environmental Factors}
For environmental factors, this study focuses on two met-ocean variables: Significant wave height, denoted by $H$, and wind speed, denoted by $v$. Significant wave height is typically considered the main determinant of offshore access, since vessels have a prescribed upper safety limit, denoted by $H^*$, above which they cannot safely carry out offshore missions. Wind speed, on the other hand, can 
compromise the safe navigation of the vessel, complicate berthing, mooring, transfer, and lifting operations, as well as limit areal modes of transport. We denote the wind speed safety limit as $v^*$. 
In this study, data about $H$ and $v$ are obtained from two main sources: (\textit{i}) A set of observational data collected using thirteen buoys in the Northeastern United States; and (\textit{ii}) A set of met-ocean data extracted from two numerical models. In what follows, we describe these datasets and their relevance to the offshore wind energy areas off the U.S. East Coast. All observational and numerical datasets used in this study are available in the public domain, and the references to access these datasets are cited in the appropriate context.
\subsubsection{Observational Data}
\begin{figure}
    \centering
    \includegraphics[width=1\linewidth]{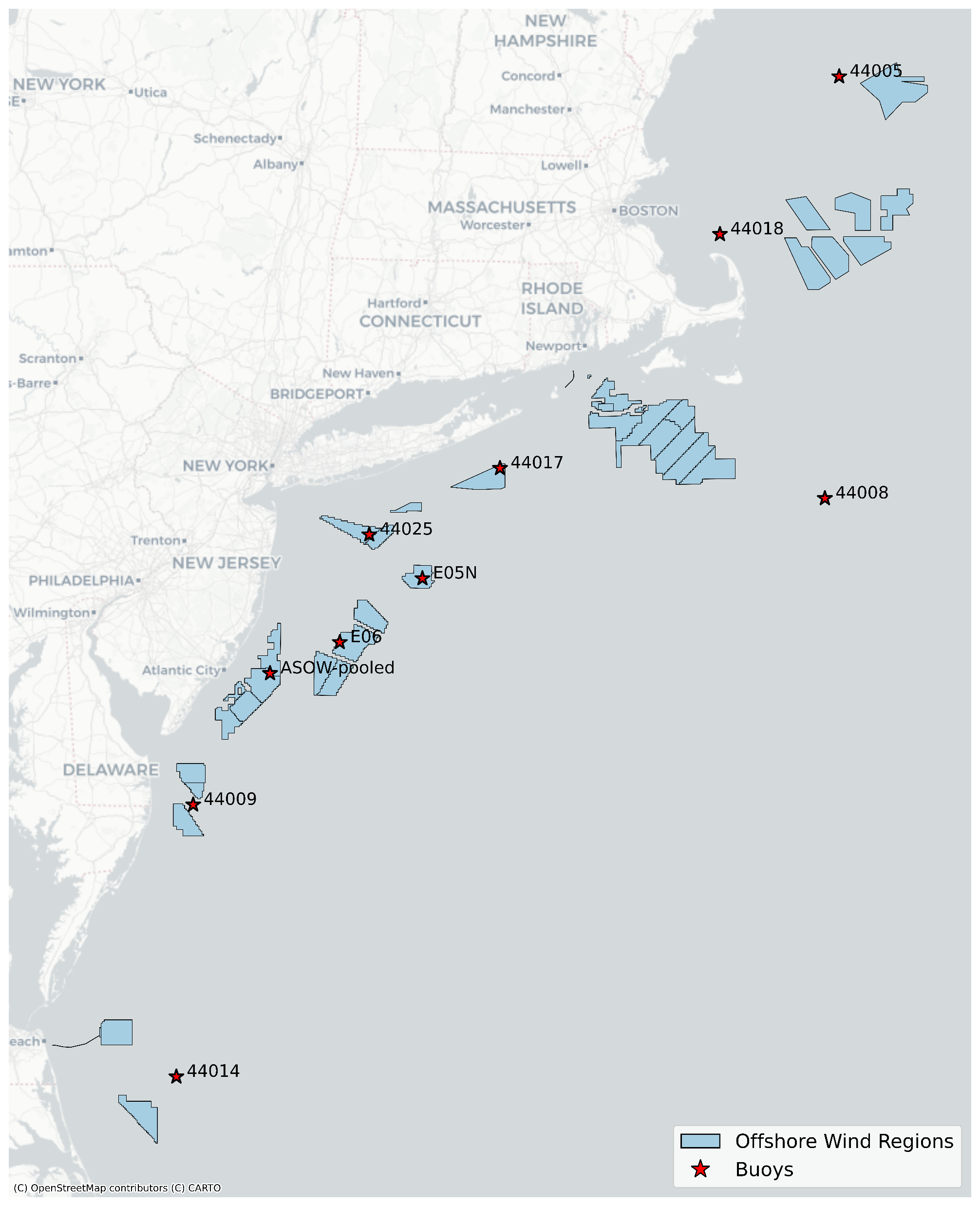}
    \caption{
        A geographical map showing the locations of the buoys (red stars) on top of the offshore wind regions off the U.S. East Coast (blue polygons), as adapted from the Bureau of Ocean Energy Management (BOEM) as of August, 2025\cite{boemmap}. There were originally four proximate ASOW buoys, for which the data have been pooled (for increased data coverage) into a central location, depicted as ASOW-pooled. More details about the observational data are in  
        Table~\ref{tab:buoy_data}.
    }
    \label{fig:map}
\end{figure}

Five years of observational data, from January 2019 to December 2023, are collected from three main sources: (\textit{i}) The National Data Buoy Center (NDBC) maintained by The National Oceanic and Atmospheric Administration (NOAA)\cite{noaa} (a total of 7 buoys), providing wind data at a height of 3 meters ($m$) as well as wave-related data; (\textit{ii}) ASOW (Atlantic Shores Offshore Wind) buoys\cite{asowdata} (a total of 4 buoys), providing $10$-$m$ wind data, as well as wave data; and (\textit{iii}) New York State Energy Research and Development Authority (NYSERDA)-supported buoys\cite{nyserdameasurement} (a total of 2 buoys), 
providing $10$-$m$ wind data, as well as wave data. Figure \ref{fig:map} shows the locations of the buoys, on top of the offshore wind energy areas off the U.S. East Coast.

Note that an even larger number of buoys in this region was initially considered, but some buoys were eliminated due to severe data missingness or insufficient temporal coverage. Additionally, considering that the ASOW buoys are in close proximity to one another, their data have been pooled for increased data coverage into a central location, which we call ASOW-pooled. Data pooling was implemented by averaging the four ASOW datasets while excluding null entries: if only one buoy reported a value at a given timestamp, that value was used; if more than one buoy reported a value, the average of all non-null values was recorded as the pooled entry. In total, five-year data from 10 different locations were selected after the pooling process was complete, starting from January 2019 to December 2023. 
The data from all three sources were recorded in either 10-minute or 1-hour intervals. All 10-minute data were processed into hourly averages. Additional data pre-processing was carried out to discard erroneous or physically implausible values to ensure the quality of the input met-ocean datasets. 


Table \ref{tab:buoy_data} details the coordinates and coverage for each buoy. Each dataset has a different time coverage, 
and several streaks of missing values. The NOAA buoys were the most complete with long coverage windows and contained near-sea-level wind and wave data. The observations from all buoys can be naturally regarded as spatial-temporal data, so we denote by $H_{it}$ and $v_{it}$ the values of the significant wave height and wind speed observed at location $i$ and time $t$, respectively, with, $i \in \{1, \hdots, I\}$ and  $t \in \{t_i^o, \hdots, t_i^e\}$, such that $I = 10$ locations, whereas $t_i^o$ and $t_i^e$ are the start and end times of the $i$th buoy's data. 
\begin{table}
    \centering
    \begin{tabular}{|c|c|c|c|c|c|c|}
    \hline
\textbf{Buoy} & \textbf{Latitude} & \textbf{Longitude} & \textbf{Variable} & \textbf{Start} & \textbf{End} & \textbf{\% Missing} \\
\hline
\multirow{2}{*}{44005} & \multirow{2}{*}{$43.20$} & \multirow{2}{*}{-$69.13$} & $H$ & 2019-01-04 18:00 & 2023-12-31 23:00 & 21.0\% (21.2\%*) \\
\cline{4-7}
 & & & $v$ & 2019-06-04 07:00 & 2023-12-31 23:00 & 24.5\% (30.9\%*) \\
\hline
\multirow{2}{*}{44008} & \multirow{2}{*}{$40.50$} & \multirow{2}{*}{-$69.25$} & $H$ & 2019-05-10 14:00 & 2023-12-31 23:00 & 1.5\% (8.5\%*) \\
\cline{4-7}
 & & & $v$ & 2019-05-10 14:00 & 2023-12-31 23:00 & 0.7\% (7.7\%*) \\
\hline
\multirow{2}{*}{44009} & \multirow{2}{*}{$38.46$} & \multirow{2}{*}{-$74.69$} & $H$ & 2019-01-01 00:00 & 2023-12-31 23:00 & 4.4\% (4.4\%*) \\
\cline{4-7}
 & & & $v$ & 2019-05-07 18:00 & 2023-12-31 23:00 & 25.7\% (30.9\%*) \\
\hline
\multirow{2}{*}{44014} & \multirow{2}{*}{$36.60$} & \multirow{2}{*}{-$74.84$} & $H$ & 2019-01-01 00:00 & 2023-12-31 23:00 & 20.7\% (20.7\%*) \\
\cline{4-7}
 & & & $v$ & 2019-01-01 00:00 & 2023-12-31 23:00 & 18.4\% (18.4\%*) \\
\hline
\multirow{2}{*}{44017} & \multirow{2}{*}{$40.69$} & \multirow{2}{*}{-$72.05$} & $H$ & 2019-01-01 00:00 & 2023-01-21 13:00 & 22.4\% (37.0\%*) \\
\cline{4-7}
 & & & $v$ & 2019-01-01 00:00 & 2023-01-21 13:00 & 20.8\% (35.8\%*) \\
\hline
\multirow{2}{*}{44018} & \multirow{2}{*}{$42.20$} & \multirow{2}{*}{-$70.15$} & $H$ & 2019-01-01 00:00 & 2023-12-31 23:00 & 26.7\% (26.7\%*) \\
\cline{4-7}
 & & & $v$ & 2019-01-01 00:00 & 2023-12-31 23:00 & 26.0\% (26.0\%*) \\
\hline
\multirow{2}{*}{44025} & \multirow{2}{*}{$40.26$} & \multirow{2}{*}{-$73.18$} & $H$ & 2019-01-01 00:00 & 2023-12-18 12:00 & 1.4\% (2.1\%*) \\
\cline{4-7}
 & & & $v$ & 2019-01-01 00:00 & 2023-09-04 19:00 & 2.6\% (8.9\%*) \\
\hline
\multirow{2}{*}{E05N} & \multirow{2}{*}{$39.97$} & \multirow{2}{*}{-$72.72$} & $H$ & 2019-08-12 00:00 & 2021-09-19 23:00 & 0.2\% (57.9\%*) \\
\cline{4-7}
 & & & $v$ & 2019-09-01 00:00 & 2021-09-14 02:00 & 11.6\% (65.0\%*) \\
\hline
\multirow{2}{*}{E06} & \multirow{2}{*}{$39.55$} & \multirow{2}{*}{-$73.43$} & $H$ & 2019-09-04 00:00 & 2022-03-27 23:00 & 6.0\% (51.8\%*) \\
\cline{4-7}
 & & & $v$ & 2019-09-04 00:00 & 2022-03-27 23:00 & 15.0\% (55.3\%*) \\
\hline
\multirow{2}{*}{ASOW-1} & \multirow{2}{*}{$39.31$} & \multirow{2}{*}{-$74.11$} & $H$ & 2021-05-17 01:00 & 2023-12-31 23:00 & 41.8\% (69.4\%*) \\
\cline{4-7}
 & & & $v$ & 2021-10-07 15:00 & 2023-12-31 23:00 & 42.9\% (74.5\%*) \\
\hline
\multirow{2}{*}{ASOW-2} & \multirow{2}{*}{$39.57$} & \multirow{2}{*}{-$73.97$} & $H$ & 2022-08-14 01:00 & 2023-10-23 20:00 & 13.2\% (79.3\%*) \\
\cline{4-7}
 & & & $v$ & 2022-08-14 01:00 & 2023-10-23 20:00 & 13.2\% (79.3\%*) \\
\hline
\multirow{2}{*}{ASOW-4} & \multirow{2}{*}{$39.20$} & \multirow{2}{*}{-$74.08$} & $H$ & 2021-05-14 17:00 & 2022-06-14 05:00 & 68.1\% (93.1\%*) \\
\cline{4-7}
 & & & $v$ & 2021-05-14 17:00 & 2022-06-14 05:00 & 68.1\% (93.1\%*) \\
\hline
\multirow{2}{*}{ASOW-6} & \multirow{2}{*}{$39.27$} & \multirow{2}{*}{-$73.94$} & $H$ & 2020-02-26 14:00 & 2021-05-14 13:00 & 47.2\% (87.2\%*) \\
\cline{4-7}
 & & & $v$ & 2020-02-26 14:00 & 2021-05-14 13:00 & 46.9\% (87.1\%*) \\
\hline
\multirow{2}{*}{ASOW-pooled} & \multirow{2}{*}{$39.34$} & \multirow{2}{*}{-$74.03$} & $H$ & 2020-02-26 14:00 & 2023-12-31 23:00 & \cellcolor{yellow!20} 32.3\% (47.9\%*) \\
\cline{4-7}
 &  &  & $v$ & 2020-02-26 14:00 & 2023-12-31 23:00 & \cellcolor{yellow!20} 32.8\% (48.3\%*) \\
\hline
\end{tabular}
\caption{Description of the observational data used in this study from thirteen different buoys, including the location, time coverage, and percentage of missing values for wave height ($H$) and wind speed ($v$). In the last column, the values preceding the parentheses denote the percentage of missing values over the available data window (between the start and end dates), whereas values in the parentheses denote the percentage of missing values over the full five-year range, from January 2019 to December 2023. In the bottom row, the yellow cells indicate the data missingness values after pooling the four ASOW buoys, which are significantly lower when compared to the four ASOW locations exclusively over the five-year range.}
\label{tab:buoy_data}
\end{table}
\FloatBarrier

\subsubsection{Numerical Model Data}
Numerical model data are sourced from NOAA’s Global Forecast System (GFS)\cite{gfs} and the GFS-driven WaveWatch III wave model\cite{wavewatch} for wind and wave data, respectively. Hereinafter, these will be collectively referred to as the \textit{numerical model data} or \textit{numerical model outputs}. These models operate on a 0.5° x 0.5° and 0.25° x 0.25° grid resolution, respectively, and provide hourly data for both wind speed and significant wave height. Numerical model outputs 
are all publicly available \cite{gfs,nomads_ncep_noaa}. 
Grid points used in this study are selected to spatially match each buoy location as closely as possible. At each buoy location, four surrounding numerical model grid points (north, south, east, and west) were extracted. After which, Delaunay Triangulation was used to interpolate the model value at the precise buoy location. This approach helps reduce spatial bias inherent in assigning the nearest grid point alone. 
We denote by $\tilde{H}_{it}$ and $\tilde{v}_{it}$ as the numerical model output of the significant wave height and wind speed at location $i$ and time $t$, respectively.   

\begin{figure}[h]
    \centering
    \includegraphics[width=\textwidth]{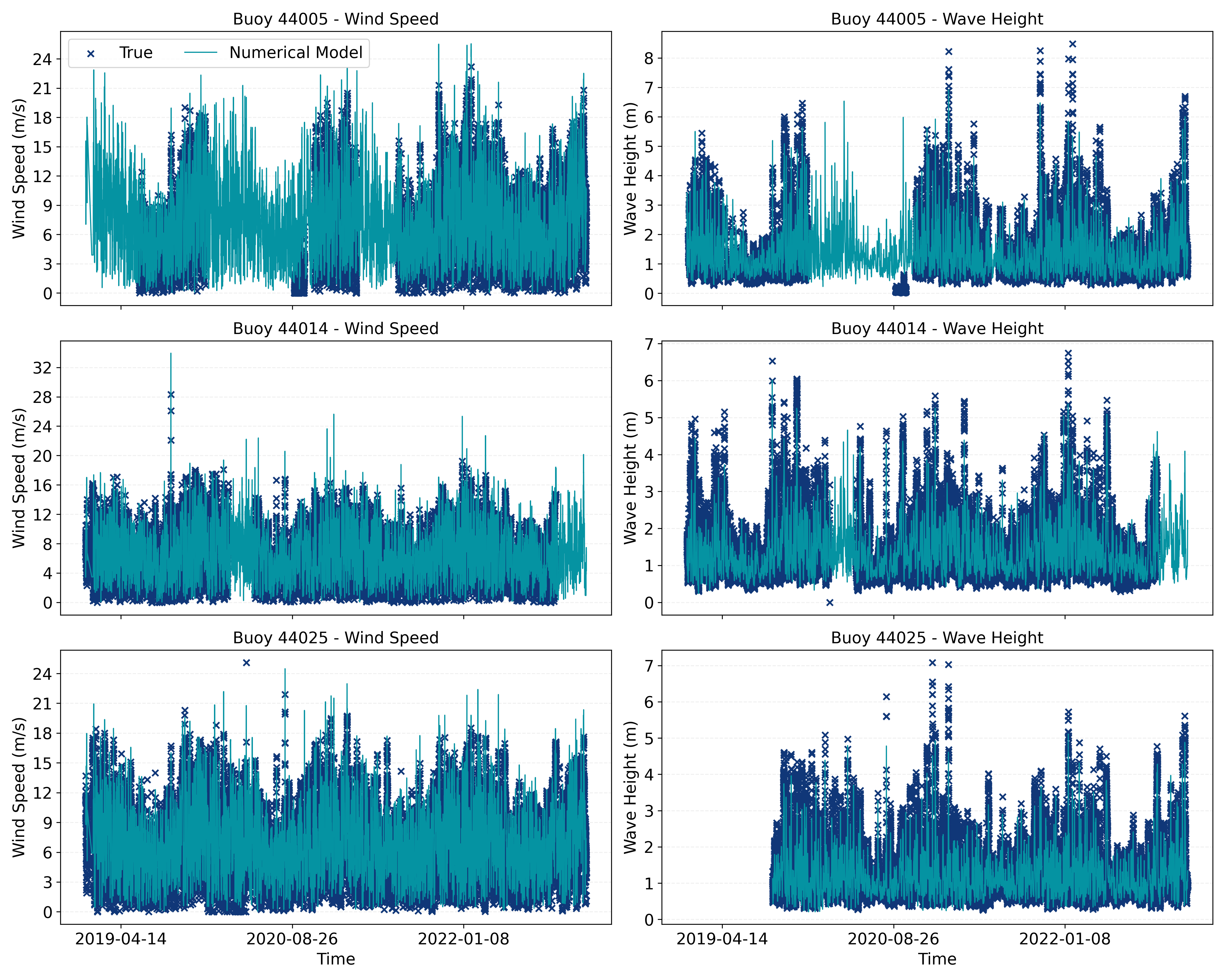}
    \caption{
        Five-year time series of wind speed (left) and significant wave height (right) from the GFS and the GFS-driven Wavewatch III, respectively (cyan lines), and buoy observations (navy cross markers) at three NOAA buoys, 44005, 44014, and 44025, from January 2019 to December 2023. While there are instances where the numerical model is unavailable, coverage is far more complete and consistent relative to observational data. 
    }
    \label{fig:timeseries_full}
\end{figure}
\FloatBarrier
Figure~\ref{fig:timeseries_full} shows the time series of numerical and observational data for wind speed (left) and significant wave height (right) at three buoy locations: 44005, 44014, and 44025. It clearly highlights the completeness of numerical model outputs compared to the substantial gaps in the observational data over the five-year period. While the numerical model outputs provide consistent temporal coverage across all sites, the observational data records show large stretches of missing data. 

\begin{figure}
    \centering
    \includegraphics[width=\textwidth]{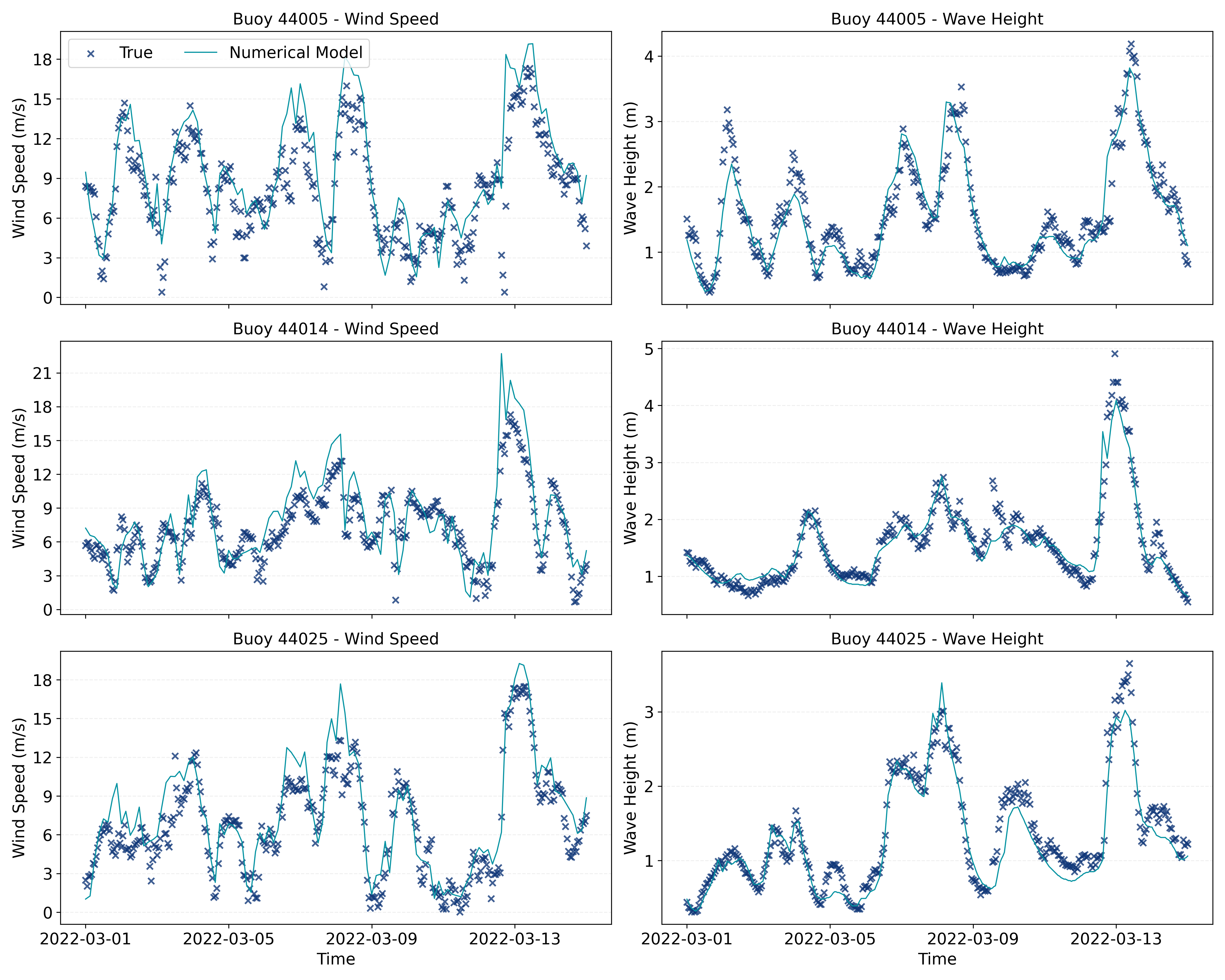}
    \caption{
        Zoomed-in view of numerical model outputs (cyan lines) versus observational data (blue cross markers) for approximately two weeks in March 2022. Despite the overall alignment, noticeable under- and over-estimation biases can be observed especially for severe met-ocean conditions which are the most relevant for the accessibility assessment in offshore operations. 
    }
    \label{fig:timeseries_zoomed}
\end{figure}

We also observe that numerical model outputs, while generally aligning with observational data, tend to exhibit noticeable biases. This is even more obvious in Figure~\ref{fig:timeseries_zoomed} which offers a zoomed-in view of the numerical model outputs over a select short time frame (about two weeks). Despite the overall alignment in both trend and magnitude, considerable under- and over-estimation biases in estimating met-ocean conditions are observable\cite{gilbert2021probabilistic,ye2024airu}. These biases appear to be especially pronounced during severe met-ocean conditions, which are most relevant for offshore accessibility considerations as they are at or near the critical safety thresholds of offshore operations. 
This trade-off between data abundance and fidelity motivates the need for a statistical treatment, as the one proposed in Section~3, which integrates both data sources (namely, the observational and numerical model data) to produce reliable, high-resolution estimates of accessibility that would be practically unattainable using observational or numerical model data alone.

\subsection{Operational Factors}
Two sets of operational parameters are considered in this study. The first operational parameter is the duration of the offshore mission to be carried out. This is defined via a minimum weather window length, denoted hereinafter as $\zeta$, which is required to safely complete this offshore mission. For example, if $\zeta = 12$ hours, then the operator needs to schedule this mission at a time when at least twelve hours of sustained access to the offshore asset is expected. The second parameter is the safety limit(s) of the met-ocean environment, which is represented by a vector of upper bounds on the met-ocean variable(s) considered, denoted in this study as  $H^*$ for significant wave height, and $v^*$ for wind speed. 

The ability to access an offshore asset is dependent on the combination of $\zeta$, $H^*$, and $v^*$, which are in turn informed by the type of offshore mission and transport system employed. In this study, we use $H^* \in [1.5 m, 4.5 m]$ in $0.25$-$m$ increments. The lower and upper bounds of this interval are chosen to reflect standard safety limits from common crew transfer vessels (CTVs) to large service operation vessels (SOVs) with advanced access systems\cite{rowell2022does}. The safety limit for wind speed is set at $12$ $ms^{-1}$, that is, $v^* = 12$ $ms^{-1}$, which has been reported in various sources as an upper bound for {aerial} transport\cite{petros2021,petros2022,turk2010offshore}. We also select various values for $\zeta$, reflecting different types of maintenance actions. Particularly, we choose $\zeta \in \{2, 6, 12, 18, 24\}$ hours, reflecting different levels of mission severity, from minor to more complex repairs 
\cite{dowell2013analysis,rowell2022does}. 

Additionally, information about the vessel's planned travel route and operational profile is typically available to logistical planners for standard offshore operations. An operational profile is a breakdown of the list of activities to be carried out by a vessel over the duration of its voyage (e.g., transiting, idling, loading/unloading), the percentage of time spent on each of these activities, as well as the travel speed and fuel consumption during each of these activities. This information, as we present later in Section 3.3., will be key inputs in defining serviceability as an operational, logistics-aware metric.

\section{Methods}
We propose a two-phase methodological framework. \textit{First}, we propose a time-series regression model which leverages both observational and numerical model data. 
This modeling approach accounts for seasonal and temporal trends, while correcting the biases of numerical model outputs, thereby producing high-resolution time series of met-ocean conditions for use in subsequent logistical analysis and planning. This is presented in Section 3.1. \textit{Second}, we use the statistical model outputs, namely the high-resolution time series of met-ocean conditions, to evaluate operational metrics such as approachability,  accessibility, and serviceability. These metrics will be defined explicitly in Sections 3.2 and 3.3. 

\subsection{Statistical Modeling of Met-ocean Conditions: Combining Numerical and Observational Data}
\label{sec:met-ocean_model}

Observational data provides high-fidelity information about met-ocean conditions, but solely using them to estimate accessibility restricts the evaluation to time periods where measurements are available. Since these time periods only span for limited periods of time and contain fairly long streaks of missing data, this would result in misleading estimates due to missing important temporal features, such as season-to-season and year-to-year variations. A viable alternative is to use numerical model data, which are fairly abundant and do not significantly suffer from data missingness issues relative to observational data. However, as presented in Section 1, numerical models exhibit considerable biases that can largely inflate the accessibility estimation errors. 

To address these limitations, we propose a statistical framework to combine the high-fidelity (but scarce) observations with the fairly abundant (but lower-fidelity) numerical model data, for improved accessibility estimates. The first step is to extract site-specific time series comprising observational and numerical model data at periods where both sets of data are available. We then regress the observations on the numerical model data at each site using a time series regression (TSR) model. Let $\mathbf{x}_{it} = [x_{it,1}, ..., x_{it,M}]^T$ denote the vector of $M$ met-ocean variables of interest at location $i$ and time $t$. For example, if we only use wave height as the sole determinant of accessibility, then, $\mathbf{x}_{it} = \{H_{it}\}$. If we use both wave height and wind speed as determinants of accessibility, then we have $\mathbf{x}_{it} = [H_{it}, v_{it}]^T$. Consequently, $\mathbf{x}^* \in \mathbb{R}^M$ is the vector of safety limits acting, element-wise, on $\mathbf{x}_{it}$. The proposed TSR model is defined as in (\ref{eq:tsr}). 
\begin{equation}
x_{it,m} = \theta_0 + \theta_1 \tilde{x}_{i,t,m} + \theta_2 \tilde{x}_{i,t-1,m} + \theta_3 \tilde{x}_{i,t-24,m}
+ \sum_{k=1}^{K} \alpha_k s_{k,t} + \sum_{k=1}^{K} \beta_k c_{k,t}
+ \sum_{k=1}^{K} \psi_k s'_{k,t} + \sum_{k=1}^{K} \omega_k c'_{k,t} + \epsilon_{it}, \quad \forall i \in \{1, ..., I\}, \quad \forall m \in \{1, ..., M\}
\label{eq:tsr}
\end{equation}
where 
$\theta_0$, $\theta_1$, $\theta_2$, and $\theta_3$ are regression coefficients for the intercept, current numerical model value $\tilde{x}_{it,m}$, and its lagged values at $t - 1$ and $t - 24$ hours, respectively. The coefficients $\{\alpha_k\}_{k=1}^{K}$ and $\{\beta_k\}_{k=1}^{K}$ are associated with the sine and cosine terms of the first trigonometric basis (monthly cycle), while $\{\psi_k\}_{k=1}^{K}$ and $\{\omega_k\}_{k=1}^{K}$ are the coefficients for the sine and cosine terms of the second trigonometric basis (yearly cycle). The basis functions $s_{k,t}$, $c_{k,t}$, $s'_{k,t}$, and $c'_{k,t}$ are defined in Equation~(\ref{eq:trig_bases}), each capturing periodic effects over different time scales. The term $\epsilon_{i,t}$ is a zero-mean error term with its variance denoted as $\eta$.
\begin{equation}
s_{k,t} = \sin\left(\frac{2\pi k t}{720}\right), \quad
c_{k,t} = \cos\left(\frac{2\pi k t}{720}\right), \quad
s'_{k,t} = \sin\left(\frac{2\pi k t}{8760}\right), \quad
c'_{k,t} = \cos\left(\frac{2\pi k t}{8760}\right),
\label{eq:trig_bases}
\end{equation}
such that $K$ is the number of Fourier pairs. We found that $K = 8$ is a reasonable choice to balance over- and under-fitting, {as assessed using the mean absolute error on an independent year-long validation set.} 

The justification for the parametric form in (\ref{eq:tsr}) is as follows. In our exploratory data analysis, two main features were noticed: seasonal variability and significant auto-correlations. This motivated the inclusion of Fourier terms, of which the trigonometric functions can capture monthly and yearly seasonalities. Additionally, the one-hour lagged term captures short-term autocorrelations, whereas the one-day lagged term captures diurnal seasonality. 
Thus, the goal of the formulation in (\ref{eq:tsr}) is to calibrate the numerical model output, whereas the trigonometric series terms account for the remaining temporal variability that is not fully explained by the numerical model. 

We fit 20 different TSR models ($10$ sites $\times$ $2$ met-ocean variables = $20$ models). 
Let $\mathbf{H}_{i,m}$ denote the regression matrix of the TSR model at the $i$th site and $m$th met-ocean variable, as expressed in (\ref{eq:regmat}). 
\begin{equation}
\mathbf{H}_{i,m} = 	\begin{bmatrix}
1 & \tilde{x}_{i,1} & s_{1,1}   &   c_{1,1} & ....  &   s_{K,1} &   c_{K,1} \\
\vdots &\vdots&\vdots& \vdots &\vdots& \vdots & \vdots \\
1   & \tilde{x}_{i,T}  & s_{1,T} &  c_{1,T} & ...   &  s_{K,T} &   c_{K,T} \\
\end{bmatrix}.
\label{eq:regmat}
\end{equation}
Then, the output of the TSR model at any time $t$ (including where no observations are available) is denoted as $\hat{x}_{it,m}$, and is determined as in (\ref{eq:tsr_mean}). 
\begin{equation}
\hat{x}_{it,m} = \mathbf{h}_{i,m}^{new}(\mathbf{H}^T_{i,m}\mathbf{H})^{-1}\mathbf{H}^T\mathbf{x}_{i,m}, 
\label{eq:tsr_mean}
\end{equation}
where $\mathbf{h}_{i,m}^{new} = [1, \tilde{x}^{new}_{i,t}, s_{1,t}, c_{1,t}, ..., s_{K,t}, c_{K,t}]^T$, such that $\tilde{x}^{new}_{i,t}$ is the numerical model output at time $t$, whereas $\mathbf{x}_{i,m} = \{x_{it,m}\}_{t=1}^{T}$.

\FloatBarrier
\subsection{Approachability and Accessibility}
Having established the TSR models, we use them to make predictions for the multi-year met-ocean conditions at each location. Those are then used to compute important operational metrics for each site. 
Approachability is defined as the fraction of time that the met-ocean conditions are below their prescribed safety limits. 
We denote the approachability at the $i$th site as $R_i(\mathbf{x}^*)$ and express it mathematically as in (\ref{eq:approach}). 
\begin{equation}
    R_i(\mathbf{x}^*) = \frac{1}{T}\sum_{t = 1}^{T}
    \bigg[\prod_{m=1}^{M} \mathbbm{1}(x_{i,t,m} < x^*_m)\bigg],
    \label{eq:approach}
\end{equation}
where $T$ is the total number of data points used to evaluate approachability, and $\mathbbm{1}(\cdot)$ is an indicator function, such that $\mathbbm{1}(\mathcal{C}) = 1$ if the condition $\mathcal{C}$ is realized, and $0$ otherwise. If the expression in (\ref{eq:approach}) is evaluated using numerical model outputs instead of observational data, then $\mathbf{x}_{it}$ would be replaced by its numerical model estimate, $\tilde{\mathbf{x}}_{it}$. 

Albeit a useful metric, $R_i(\mathbf{x}^*)$ is not sufficient to inform offshore construction and O\&M planning, because offshore operations such as maintenance tasks require sustained access to the offshore asset for a sufficiently long period of time. Hence, accessibility is the metric that reflects this additional requirement by searching for consecutively approachable met-ocean behavior for the minimum duration of the weather window, $\zeta$. Unlike approachability, estimating accessibility considers the intersection of timesteps and met-ocean covariates, where a location is only considered accessible at a mission start time $t^*$ if the safety limit for each met-ocean variable is not exceeded for the entire duration of the mission. This can be mathematically expressed as in (\ref{eq:access}). 
\begin{equation}
    A_i(\mathbf{x}^*, \zeta) = \frac{1}{T-\zeta+1}
    \sum_{t^* = 1}^{T-\zeta+1} 
    \bigg
    [\prod_{m=1}^{M}\prod_{t=t^*}^{t^*+\zeta-1} \mathbbm{1}(x_{i,t,m} < x^*_m)\bigg]
    \label{eq:access}
\end{equation}
Similar to approachability, if the expression in (\ref{eq:access}) is evaluated using numerical model outputs instead of observational data, then $\mathbf{x}_{it}$ would be replaced by its numerical model estimate, $\tilde{\mathbf{x}}_{it}$.  
\FloatBarrier
\subsection{Serviceability: A Proposed Operational Metric to Assess Accessibility Along a Vessel Route}

Traditional definitions of accessibility focus on a single offshore site (one spatial location). 
While this bears some operational value, it does not consider the fact that offshore operations are \textit{not} spatially static and requires a vessel to traverse between ports to offshore assets through highly dynamic, and spatially varying met-ocean conditions. As such, accessibility at a single location (be it the port or the offshore site) 
can be insufficient to realistically map the likelihood of mission success. This spatial aspect will only increase in importance as offshore wind turbines continue to be installed in deeper waters. Recent researchers acknowledged the need for developing rigorous operational metrics to estimate the accessibility along the travel path of a vessel (not just at the site or the port), especially as tow-to-shore activities become more relevant to larger-scale turbine installations\cite{mcmorland2024operations}. 

To bridge this gap, we introduce the new operational metric of \textit{serviceability}, which we define as the accessibility of the entire operational route of a vessel for the duration of a mission. Unlike accessibility, which considers persistence of weather windows at just \textit{one spatial location}, serviceability considers the changing position of a vessel (such as a CTV), including transit from port, on-site work, and return to port. 

\begin{figure}
    \centering
    \includegraphics[width=1\linewidth]{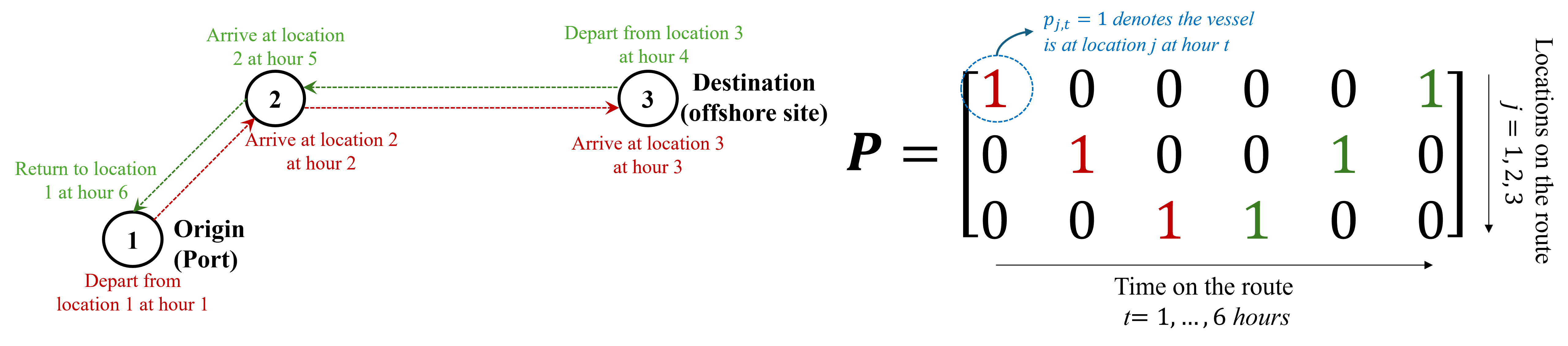}
    \caption{Left: Toy example of a vessel route, showing the vessel activities from an operational profile, mapped over time. The red text denote port-to-site voyage activities, whereas the green text denotes the return trip activities. Right: Corresponding position matrix, $\mathbf{P}$, constructed from the vessel route and operational profile information. The position matrix $\mathbf{P}$, encoding the expected location of the vessel at each time step, will be a critical input to serviceability calculations.}
    \label{fig:serviceability}
\end{figure}

Clearly, a critical input to serviceability will be information on the expected vessel route and operational profile{, listing the description and durations of activities to be carried out by a vessel over the span of its voyage}. {This information is typically available to mariners and logistical planners for standard offshore operations of any duration or complexity. The} operational profile is converted into a spatial-temporal position matrix, $\mathbf{P}$, for which each entry $p_{j,t}$ is a binary variable indicating whether the vessel is expected to be at location $j$ on the route at time $t$. Figure \ref{fig:serviceability} (left) shows a toy example of a route for a six-hour mission, comprising the port (location 1), an intermediate site (location 2), and the offshore site (location 3). This route information, along with the vessel operational profile, are used to construct the vessel position matrix (right), which will be used in serviceability calculations. 

We define serviceability over a route $r$ as a function of the met-ocean safety limits $\mathbf{x}^* = [x_1^*, x_2^*, ..., x_M^*]^T$, the operational window $\zeta$, and the position matrix $\mathbf{P}$, as in (\ref{eq:serviceability}):
\begin{equation}\label{eq:serviceability}
S_r(\mathbf{x}^*, \zeta, \mathbf{P}) = \frac{1}{T-\zeta+1} \sum_{t^* = 1}^{T-\zeta+1} 
\bigg[\prod_{\substack{j = 1, \cdots, J(r) \\ t = t^*, \cdots, t^*+\zeta-1 \\ p_{j,t} = 1}} \prod_{m=1}^{M} \mathbbm{1}(x_{j,t,m} < x^*_m)\bigg], \quad \textit{such that} \quad t^* + \zeta - 1 \leq T,
\end{equation}
where $x_{j,t,m}$ denotes the value of the $m$th met-ocean variable (e.g., significant wave height or wind speed) at location $j$ and time $t$. The route is discretized into $J(r)$ spatial points that could be potentially visited by the vessel that departs at a mission start time of $t^*$. 
The mathematical expression in (\ref{eq:serviceability}) enforces that all met-ocean conditions must be under their prescribed safety limits at every location within the route visited by the vessel (as encoded in the condition $p_{j,t} = 1$). A violation at any visited location or for any variable results in a zero-valued serviceability score for that mission start time. Averaging across all start mission times yields a final, averaged, route-specific serviceability score $S_r(\mathbf{x}^*, \zeta, \mathbf{P})$, which reflects the proportion of time when the full mission could be completed along this vessel route without interruption. 

It is evident from the above that serviceability is inherently a spatial-temporal, and route-specific metric as it directly utilizes the information on the position of the vessel 
and the expected met-ocean behavior at the \textit{vessel's} position over time. Thus, it provides a more realistic assessment of mission success likelihood along the parameters observed throughout the mission. This is in stark contrast to spatially-independent metrics such as accessibility and approachability, which overlook the vessel's travel path and only consider the met-ocean conditions at a single location, be it the port or the offshore site. 

\FloatBarrier
\section{Results}
Before presenting the accessibility and serviceability analysis, we first begin in Section 4.1. by demonstrating the value of the statistical met-ocean model previously presented in Section~\ref{sec:met-ocean_model}. This is then followed by the accessibility and serviceability analyses, which are presented in Sections 4.2 and 4.3. 

\subsection{Met-ocean Model Results}
To evaluate the performance of the met-ocean model, we use four-year data for training (Jan 2019 to Dec 2022), and then use the fifth year for testing (Jan 2023 - Dec 2023). We use three key metrics, evaluated on the test set: 
\textit{Coefficient of determination} ($R^2$), which reflects the model’s explanatory power; 
\textit{Root Mean Square Error} (RMSE), which quantifies the square root of the average squared deviation; 
and \textit{Bias}, which measures the mean signed error between predictions and true values. We compare the performance of the TSR model to the raw numerical model output without bias correction. 

Table~\ref{tab:model_performance1} shows the performance of the TSR model versus the raw numerical model output across all buoys in the test set, for significant wave height (left) and wind speed (right). 
Overall, the TSR model shows consistent improvements for most locations, across all metrics, and for both variables. The improvement is more significant for wind speed, where the TSR model achieves an average $R^2$ of $0.744$ compared to $0.636$ for the numerical model, further reducing the RMSE from $2.036$ to $1.720$. Additionally, the bias is nearly eliminated (–$0.030$ for the TSR model vs. $+0.847$ for the numerical model). Considerable (although smaller) improvements are also reported for significant wave height, where the TSR model achieves an $R^2$ of $0.895$ compared to a value of $0.885$ for the numerical model, and an RMSE of $0.254$ compared to a value of $0.263$. While the bias remains almost the same in absolute terms, TSR's slight over-estimation ($+0.030$) may be considered better than the numerical model's under-estimation ($-0.026$), since a conservative approach is safer than a liberal one in offshore operations, especially those involving personnel transfer. Collectively, these results suggest a more calibrated estimation of met-ocean conditions and support the adoption of the TSR met-ocean model for subsequent accessibility and serviceability analysis. Later in Section 4.3., we show that these differences in estimation performance can have a fairly significant impact on the resulting estimates of accessibility and serviceability. 
\begin{table}
\centering
\renewcommand{\arraystretch}{1.2}
\begin{tabular}{|c|ccc|ccc||ccc|ccc|}
\hline
    &   \multicolumn{6}{c||}{Significant Wave Height, $H$}   &   \multicolumn{6}{c|}{Wind Speed, $v$} \\
\hline
\textbf{Buoy} & \multicolumn{3}{c|}{\textbf{TSR (Proposed)}} & \multicolumn{3}{c||}{\textbf{Numerical}} & \multicolumn{3}{c|}{\textbf{TSR (Proposed)}} & \multicolumn{3}{c|}{\textbf{Numerical}} \\
             & $R^2$ & RMSE & Bias & $R^2$ & RMSE & Bias & $R^2$ & RMSE & Bias & $R^2$ & RMSE & Bias \\
\hline

44005 &  \cellcolor{green!25}0.893 & \cellcolor{green!25}0.281 & -0.019 
& 0.885 & 0.308 & \cellcolor{green!25}-0.033 
& \cellcolor{green!25}0.760 & \cellcolor{green!25}1.870 & \cellcolor{green!25}0.137 
& 0.652 & 2.251 & 1.003 
\\

44008 & \cellcolor{green!25}0.934 & \cellcolor{green!25}0.266 & \cellcolor{green!25}0.005 
& 0.929 & 0.277 & -0.033 
& \cellcolor{green!25}0.778 & \cellcolor{green!25}1.653 & \cellcolor{green!25}0.046 
& 0.664 & 2.034 & 0.916 
\\

44009 & 0.864 & 0.226 & -0.077 
& 0.867 & 0.228 & -0.082 
& \cellcolor{green!25}0.773 & \cellcolor{green!25}1.534 & \cellcolor{green!25}-0.216 
& 0.734 & 1.661 & 0.478 
\\

44014 & 0.910 & 0.241 & 0.071 
& \cellcolor{green!25}0.915 & \cellcolor{green!25}0.234 & \cellcolor{green!25}0.068 
& \cellcolor{green!25}0.697 & \cellcolor{green!25}1.746 & \cellcolor{green!25}0.063 
& 0.567 & 2.085 & 0.839 
\\

44017 &  \cellcolor{green!25}0.903 & \cellcolor{green!25}0.221 & 0.054 
& 0.897 & 0.226 & -0.069 
& \cellcolor{green!25}0.735 & \cellcolor{green!25}1.690 & \cellcolor{green!25}0.027 
& 0.592 & 2.097 & 0.934 
\\

44018 & 0.875 & \cellcolor{green!25}0.261 & -0.100 
& 0.872 & 0.266 & -0.098 
& \cellcolor{green!25}0.753 & \cellcolor{green!25}1.741 & \cellcolor{green!25}-0.027 
& 0.660 & 2.040 & 0.818 
\\

44025 & 0.883 & 0.216 & -0.044 
& 0.888 & 0.212 & -0.052 
& \cellcolor{green!25}0.730 & \cellcolor{green!25}1.730 & \cellcolor{green!25}-0.204 
& 0.657 & 1.951 & 0.649 
\\

E05N &  \cellcolor{green!25}0.889 & \cellcolor{green!25}0.239 & 0.055 
& 0.874 & 0.253 & \cellcolor{green!25}0.048 
& \cellcolor{green!25}0.727 & \cellcolor{green!25}1.683 & \cellcolor{green!25}0.122 
& 0.587 & 2.070 & 0.968 
\\

E06 & \cellcolor{green!25}0.875 & \cellcolor{green!25}0.247 & 0.038 
& 0.858 & 0.263 & \cellcolor{green!25}0.010 
& \cellcolor{green!25}0.790 & \cellcolor{green!25}1.779 & \cellcolor{green!25}0.454 
& 0.590 & 2.281 & 1.495
\\

ASOW-pooled & 0.782 & 0.276 & 0.122 
& \cellcolor{green!25}0.824 & \cellcolor{green!25}0.247 & \cellcolor{green!25}0.109 
& \cellcolor{green!25}0.697 & \cellcolor{green!25}1.774 & \cellcolor{green!25}-0.102 
& 0.657 & 1.890 & 0.370 
\\
\hline
\textbf{Average} & \cellcolor{green!25}0.895 & \cellcolor{green!25}0.254 & 0.030 
& 0.885 & 0.263 & \cellcolor{green!25}-0.026 
& \cellcolor{green!25}0.744 & \cellcolor{green!25}1.72 & \cellcolor{green!25}0.030 
& 0.636 & 2.036 & 0.847 
\\
\hline
\end{tabular}
\caption{Model performance evaluation comparing $R^2$, RMSE, and Bias for TSR vs. numerical models across all buoys, for significant wave height (left) and wind speed (right). Green cells indicate best performance per metric.}
\label{tab:model_performance1}
\end{table}

\begin{figure}
    \centering
    \includegraphics[width=\textwidth]{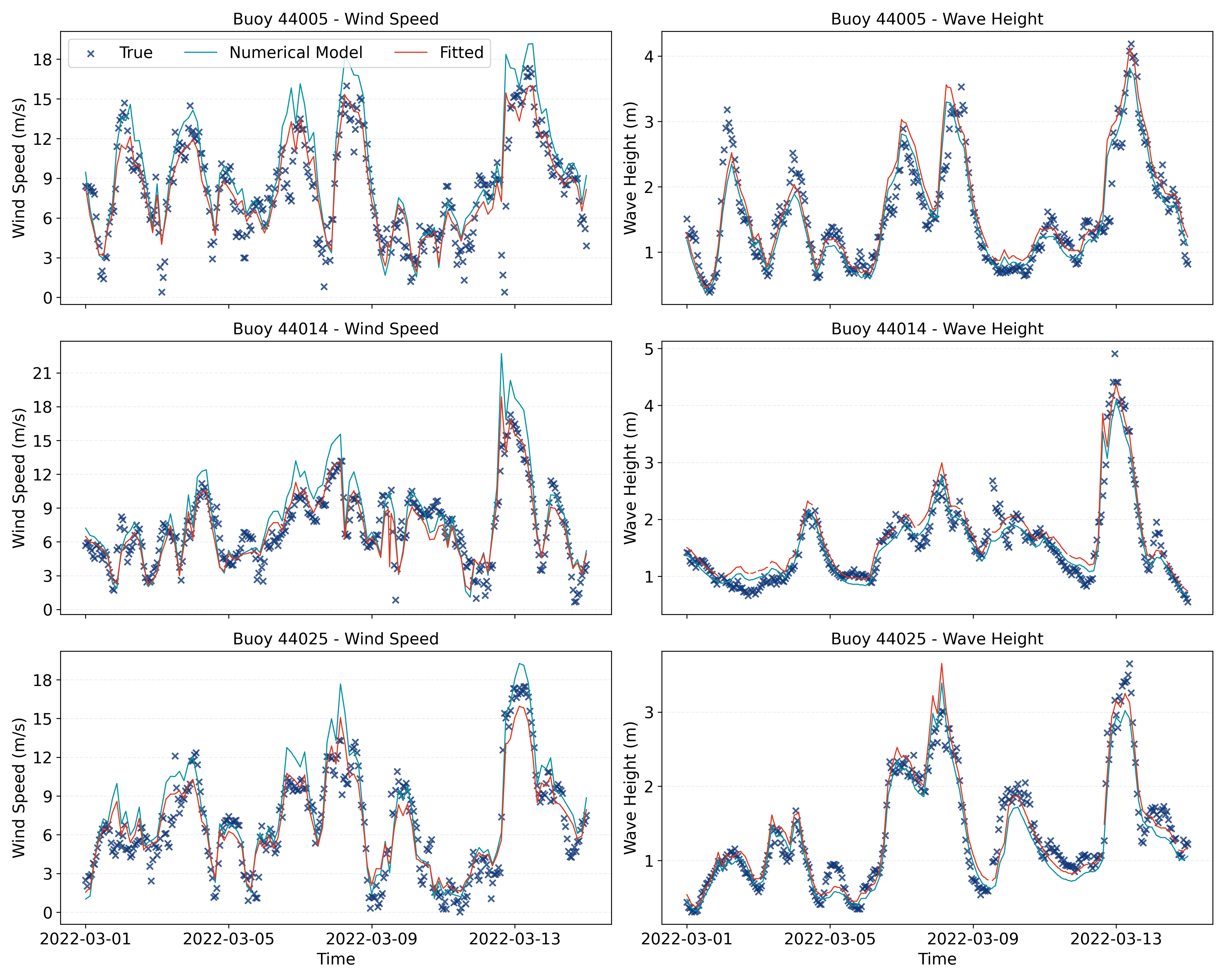}
    \caption{
        Zoomed time series comparison of numerical model (teal lines), true observations (navy markers), and the TSR model fit (red lines) for wind speed (left) and significant wave height (right), across buoys 44005, 44014, and 44025 during March 2022. The fitted TSR model tracks observed values more closely than the raw numerical model output, especially during peaks and sudden transitions, indicating improved predictive skill relative to the base numerical model.
    }
    \label{fig:tsr_zoomed_fitted}
\end{figure}

Figure~\ref{fig:tsr_zoomed_fitted} shows the time series data from three different buoys on the Atlantic coast: 44005, 44014, and 44025, for wind speed (left) and significant wave height (right). This is the same data presented earlier in Figure~\ref{fig:timeseries_zoomed}, but this time, 
with the TSR model fit (red line). As is visually apparent, the TSR model leads to a considerable reduction in the biases of the numerical model, better reflecting the underlying met-ocean behavior, especially during severe conditions. 

\subsection{Approachability, Accessibility, and Serviceability Assessment}
We now turn our focus to the accessibility and serviceability analysis. In Section 4.2.1., we conduct approachability and accessibility analysis on all buoys {shown in the geographical map of} Figure \ref{fig:map}. This is followed in Section 4.2.2. by a serviceability assessment of a sample realistic route on the U.S. East Coast. In both analyses, we conduct a multivariate assessment using both significant wave height and wind speed variables. 
Given that the safety threshold for wind speed is fairly liberal ($v^{*} = 12 m/s$), there was no significant difference in most cases between the multivariate and univariate case when significant wave height as the sole met-ocean variable. As such, in most of the subsequent analyses, results from the multivariate scenario will be presented, unless otherwise noted.

\subsubsection{Approachability and Accessibility Analysis for the U.S. East Coast}

\begin{figure}
    \centering
    \includegraphics[width=\textwidth]{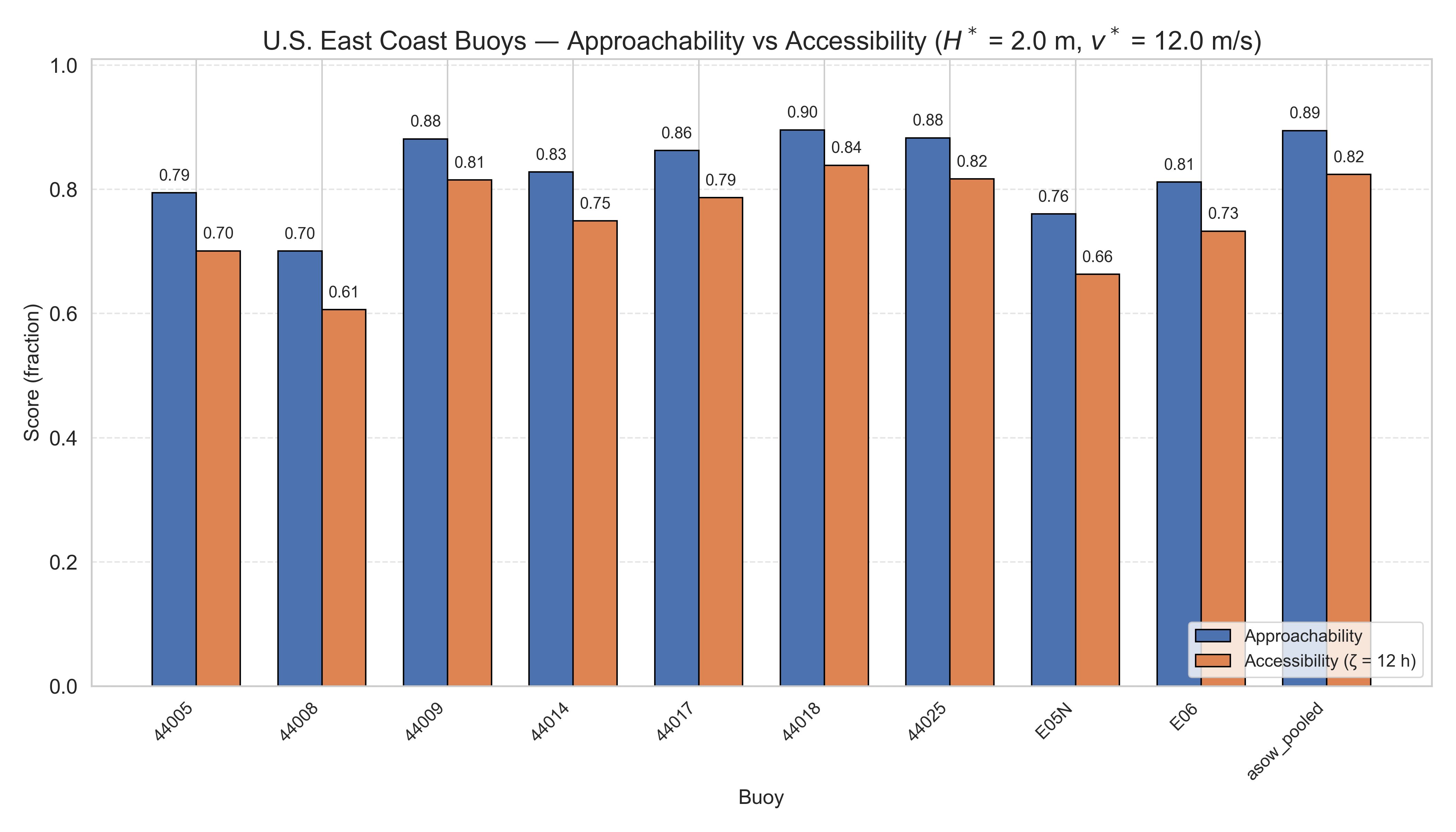}
    \caption{Comparison of approachability (blue) and accessibility (orange) for the 10 U.S. East Coast locations. The parameters used are safety limits of $H^{*} = 2.0 \,\text{m}$, $v^{*} = 12.0 \,\text{m/s}$, and the mission duration, $\zeta = 12 \,\text{h}$.}
    \label{fig:buoy_approach_access}
\end{figure}

We compare the approachability and accessibility of the $10$ locations depicted in Figure \ref{fig:map} along the U.S. East Coast. Figure~\ref{fig:buoy_approach_access} summarizes this comparison: approachability values are consistently higher than accessibility values for the same location, reflecting the stricter (and more realistic) weather window persistence requirement. In fact, approachability, on average, is consistently $5$ to $10$\% greater than accessibility. This further confirms the importance of distinguishing between these two metrics, and why the inclusion of temporal persistence, as encoded in the accessibility metric, is more relevant to offshore operations. 

\vspace{1em}

\begin{figure}
    \centering
    \includegraphics[width=\textwidth]{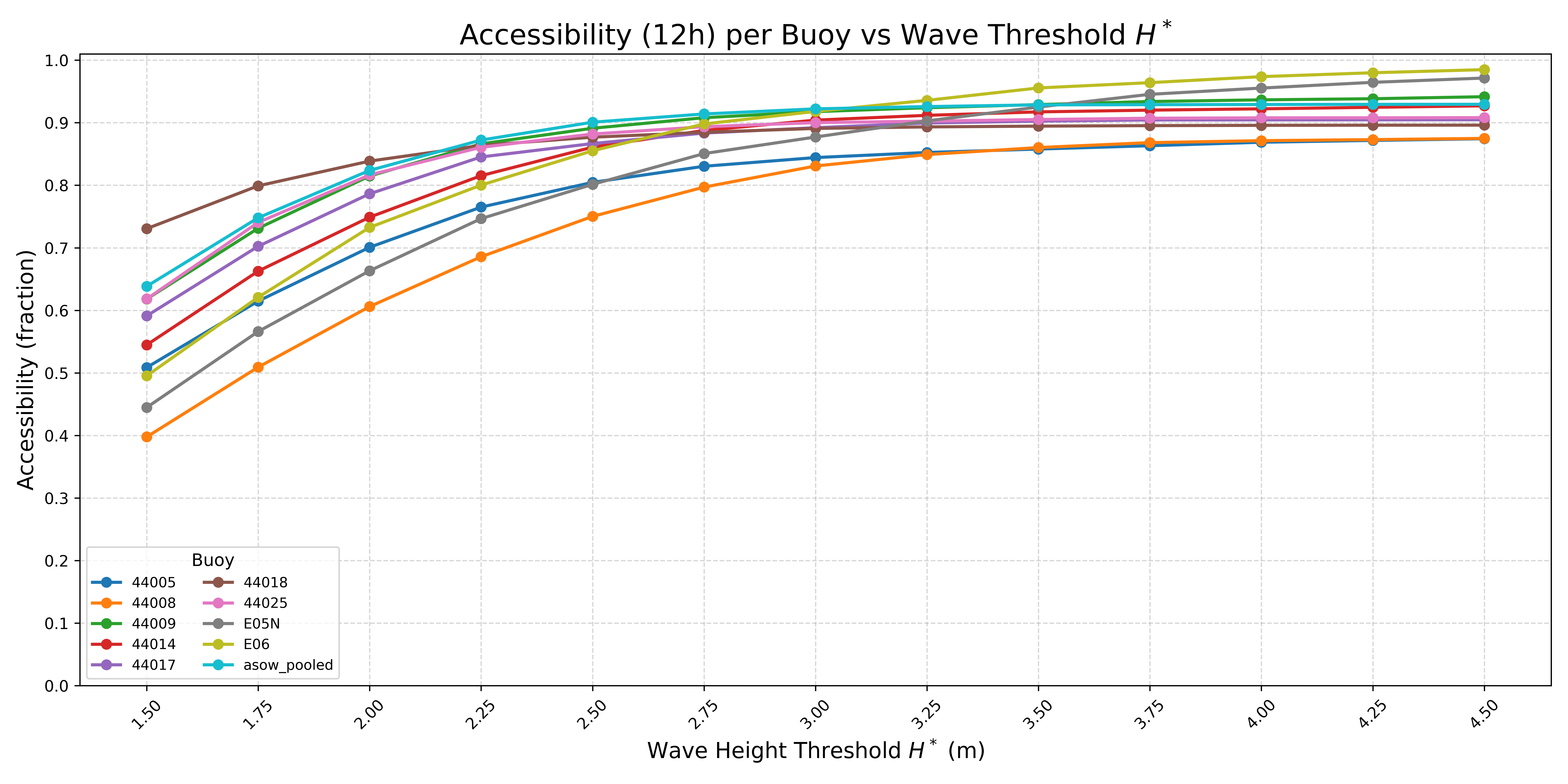}
    \caption{Accessibility for various locations (different lines) as a function of the significant wave height threshold, $H^{*}$. Higher thresholds yield higher accessibility, though spatial variability is evident. Here, a mission duration of $\zeta = 12 \,\text{h}$ is assumed.}
    \label{fig:tsr_buoy_accessibility}
\end{figure}

Figure~\ref{fig:tsr_buoy_accessibility} shows the change in accessibility across different locations (different lines) and across various wave height thresholds, $H^{*}$ (x-axis). As expected, accessibility increases monotonically as the significant wave height threshold, $H^*$ increases. The rate and magnitude of increase is location-specific. For instance, buoy 44008 has the lowest accessibility score across all wave height thresholds, demonstrating consistent, harsher conditions. This is expected considering that it is the furthest offshore location from shore, again highlighting the challenges of deeper offshore wind installations, relative to those in shallower waters. On the other hand, buoy 44018 has the largest accessibility score at the $1.5$-$m$ threshold, yet the third lowest of the ten buoys at the $4.5$-m threshold. This is indicative of a heavy-tailed distribution for the met-ocean conditions, where there is a significant portion of wave height values that fall below the $1.5$-m threshold; yet, a non-negligible number of severe environmental conditions. 

Another interesting observation in Figure~\ref{fig:tsr_buoy_accessibility} is the clear spatial differences in accessibility. 
This highlights a limitation of using accessibility to estimate the probability of successfully servicing an offshore location. For instance, if a vessel were to travel a distance comparable to that between buoy 44018 and buoy 44005 (approximately one degree to the northeast), relying on either buoy’s accessibility score in isolation would misrepresent the likelihood of favorable conditions along the full route. At the lowest threshold, the two buoys differ in accessibility by more than $20$\%, underscoring how spatial variability can mislead mission feasibility assessments when only a single-point analysis is used, and further motivating the need for a route-specific logistical metric like serviceability.  

\begin{figure}
    \centering
    \includegraphics[width=\textwidth]{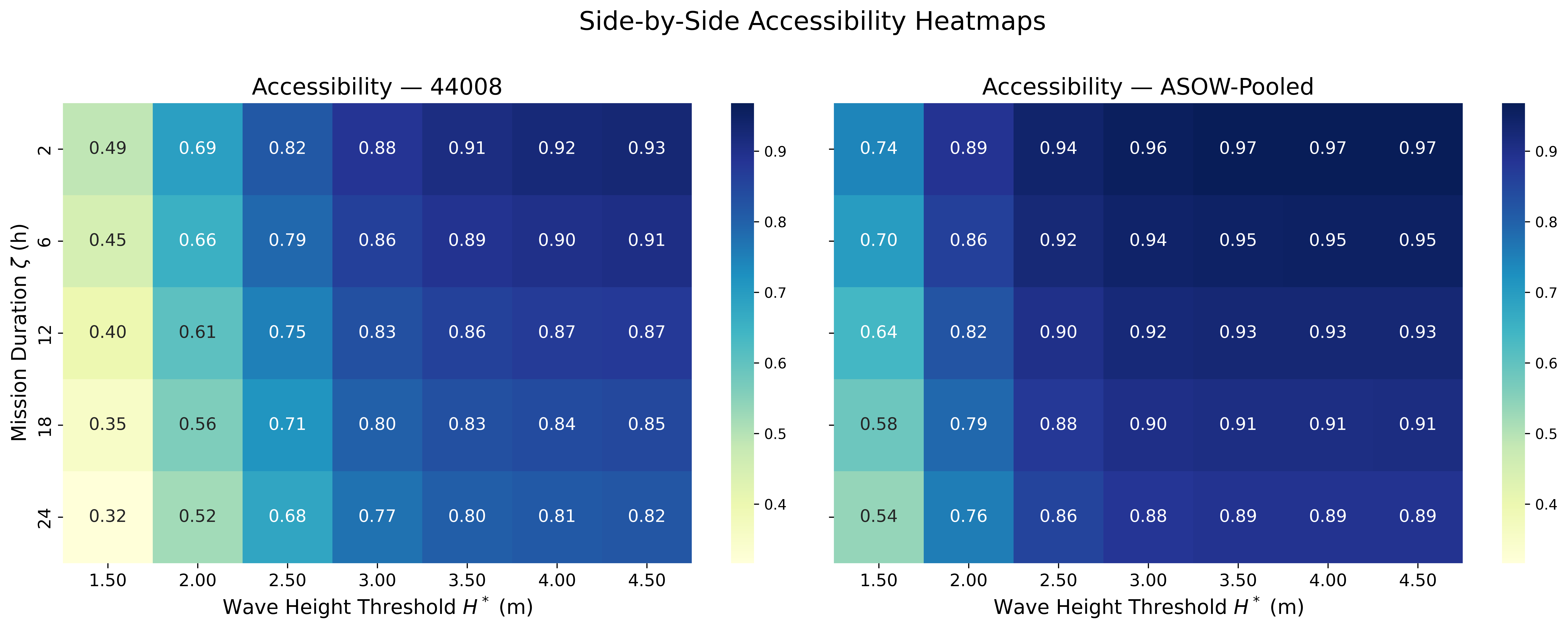}
    \caption{Side-by-side heatmaps of accessibility as a function of mission duration $\zeta$ and wave height threshold $H^{*}$. Results are shown for buoy 44008 (left) and the ASOW-Pooled dataset (right).}
    \label{fig:heatmap_44008_asow}
\end{figure}

To further elucidate the spatial differences in accessibility, Figure~\ref{fig:heatmap_44008_asow} presents side-by-side heatmaps of accessibility as a function of both mission duration $\zeta$ and wave height threshold $H^{*}$ for two locations with vastly different met-ocean conditions: buoy 44008 (farthest from shore) and the ASOW-Pooled (one of the closest to shore). The results illustrate how accessibility decreases as mission duration increases, which follows from the stricter weather window persistence requirement. However, the rate of decline differs markedly between the two cases. For buoy 44008, accessibility scores fall from $0.49$ to $0.32$ for the lowest $H^*$ threshold (a decrease of $17$\%). However, the accessibility scores for ASOW-Pooled reduce more drastically, falling from $0.74$ to $0.54$. Despite this, the trends in the data are somewhat similar, with increasing accessibility score as $H^*$ increases and $\zeta$ decreases. Yet, the importance of spatial considerations in mission success calculations is apparent, as a mission near buoy 44008 would be far less likely to yield favorable conditions at any threshold than the same mission near the ASOW buoys, suggesting a trade-off in site favorability due to proximity to shore, and hence, significantly lower logistical requirements, but potentially lower energy production levels from the offshore wind farm.  

\begin{figure}
    \centering
    \includegraphics[width=\textwidth]{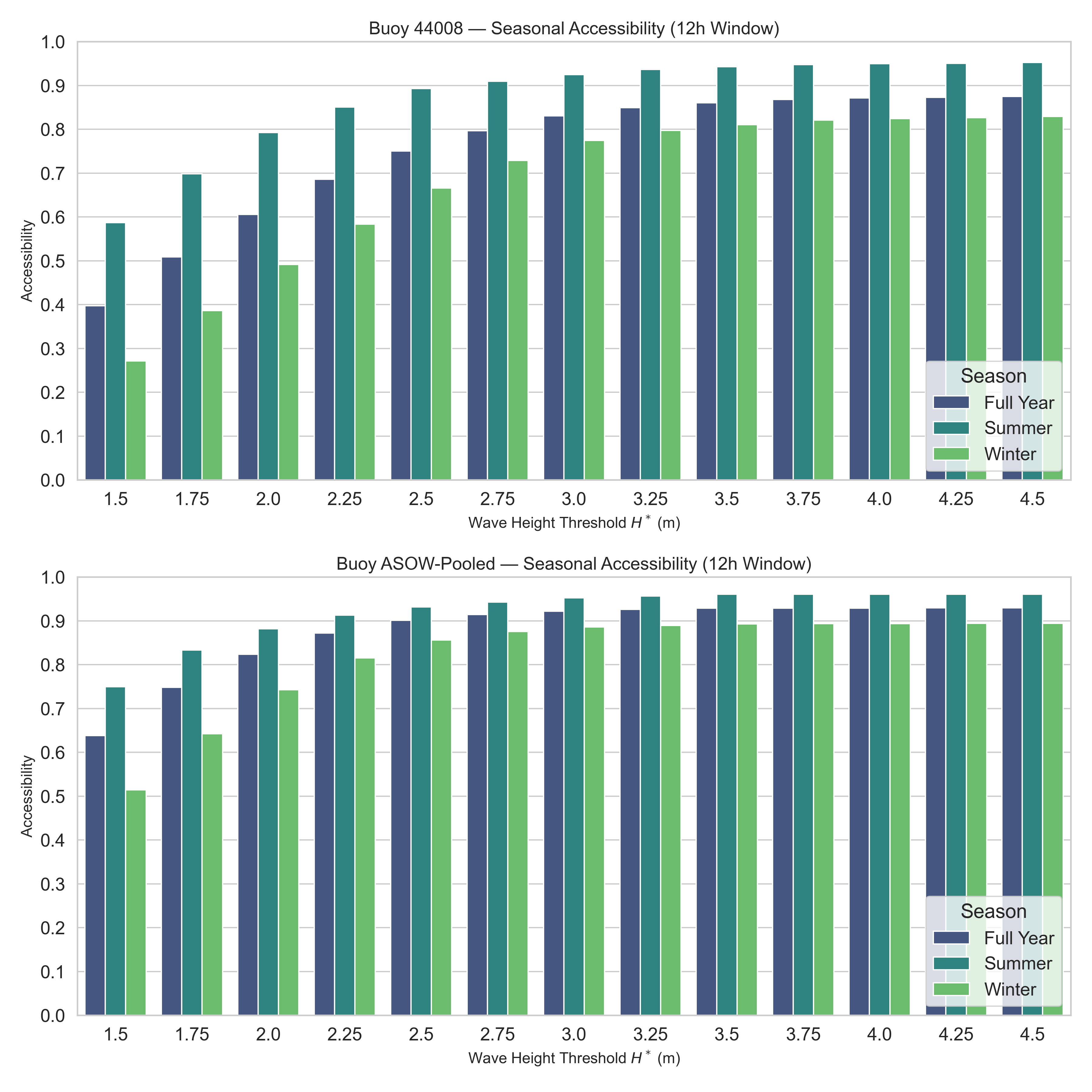}
    \caption{Seasonal accessibility as a function of significant wave height threshold $H^{*}$, shown for location 44008 (top) and ASOW-pooled (bottom). Results are presented for the full year, summer, and winter, with a mission duration of $\zeta = 12$-$h$.
    }
    \label{fig:seasonal_accessibility}
\end{figure}

Figure~\ref{fig:seasonal_accessibility} examines seasonal variations in accessibility for the same two locations: buoy 44008 (farthest from shore) and the ASOW-Pooled (one of the closest to shore). Specifically, we compare the overall accessibility scores versus those for summer and winter periods, across different wave height thresholds. The results reveal clear seasonal differences. For example, for buoy 44008 (top panel), the summer months have significantly higher accessibility scores, rising from $0.27$ in the winter to $0.59$ at the $1.5$-$m$ wave height threshold. The seasonal difference is not quite as pronounced, yet still apparent, for the ASOW-pooled location. Here, accessibility increases from $0.50$ to $0.75$ at the $1.5$-$m$ wave height threshold. In either case, there is clear temporal variations in accessibility values. This would be highly pertinent to strategic O\&M planning, as operators may choose to schedule the most important repairs and servicing actions during the summer months\cite{byon2010season, besnard2012model}.

\subsubsection{Serviceability}
The analysis in Section 4.2.1. demonstrates that accessibility varies substantially across both space and time. Spatial differences particularly highlight how reliance on a single-location estimate can mis-represent operational planning, whereas temporal variations reveal how mission duration and seasonality can critically influence the persistence of favorable conditions. As such, these findings demonstrate that, while informative, accessibility alone can be insufficient to realistically reflect the likelihood of mission success. This emphasizes the need for a more relevant logistical metric, \textit{\textbf{serviceability}}, to more precisely account for the spatio-temporal dynamics of offshore operations. 

\begin{figure}
    \centering
    \includegraphics[width=0.9\textwidth]{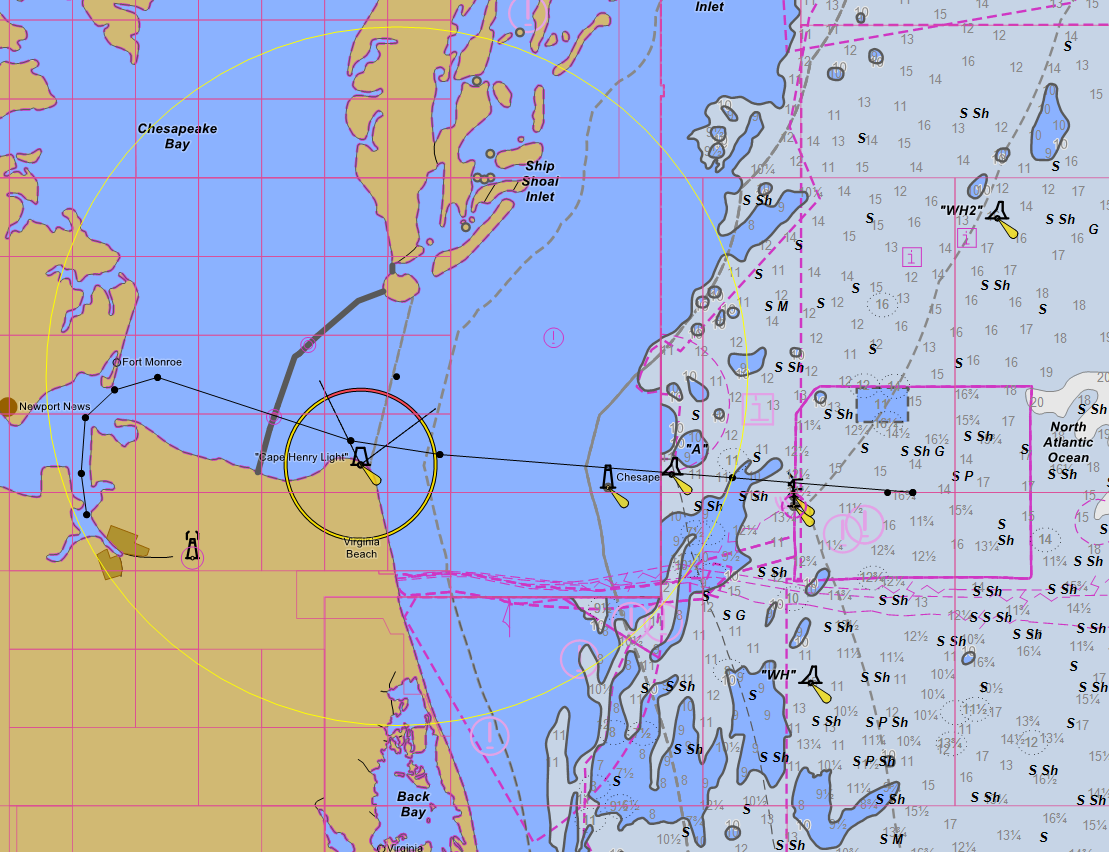}
    \caption{
Nautical chart showing a representative Mid-Atlantic Crew Transfer Vessel (CTV) service route (shown as a black solid line), departing from the Norfolk harbor area to an offshore wind location near the entrance to Chesapeake Bay. The background image is sourced from NOAA's Electronic Navigation Charts\cite{NOAA2025CHART}.}
    \label{fig:midatl_service_route}
\end{figure}

As a case study, we consider a $12$-hour round-trip mission carried out by a CTV for a specific route in the U.S. Mid-Atlantic region (MAR), {but our framework is generalizable to routes of varying durations and complexities}. The MAR region is home to several leased and planned offshore wind energy projects. Figure~\ref{fig:midatl_service_route} displays the nautical chart for the MAR, along with the specific CTV travel path (shown as black lines), as provided by our industry partners at Blue Ocean Transfers (BOT). The mission's operational profile provided by BOT dictates that the vessel departs from port for a five-hour voyage until it arrives at an offshore location near the entrance to Chesapeake Bay, where it conducts servicing activities for three hours, before returning back to port. We then use the vessel speed inputs included in the operational profile to determine the vessel's position along the route, which is discretized into six spatial locations (port, destination, and four intermediate points). Except for the destination where the vessel spends three hours, the vessel occupies each location for an hour, resulting in the following position matrix, $\mathbf{P}$. 
$$\mathbf{P} = \begin{bmatrix}
    1 0 0 0 0 0 0 0 0 0 0 0 \\
    0 1 0 0 0 0 0 0 0 0 0 1 \\
    0 0 1 0 0 0 0 0 0 0 1 0 \\
    0 0 0 1 0 0 0 0 0 1 0 0 \\
    0 0 0 0 1 0 0 0 1 0 0 0\\
    0 0 0 0 0 1 1 1 0 0 0 0\\
\end{bmatrix} $$

\begin{figure}
    \centering
    \includegraphics[width=0.8\textwidth]{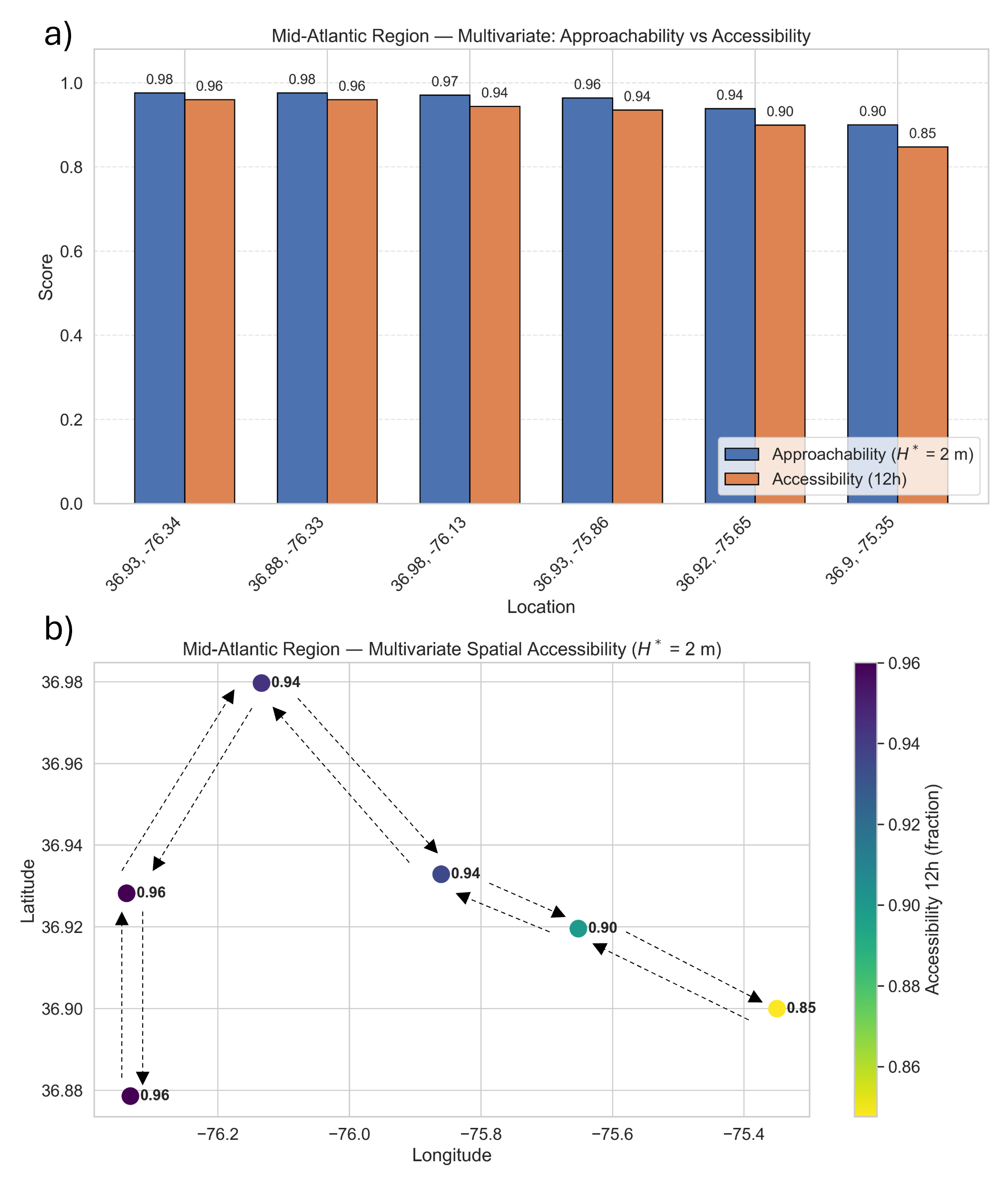}
    \caption{Panel A: Approachability and Accessibility for each location on the route. Panel B: Spatial variation in accessibility along the route. Assumptions of $H^* = 2m$, $v^* = 12 ms^{-1}$, and $\zeta = 12h$ are made. Arrows represent the MAR vessel path denoted in Figure~\ref{fig:midatl_service_route}, where the vessel moves from the bottom-leftmost position (port) to the rightmost position (wind farm) and back. }
    \label{fig:approach_access}
\end{figure}

Figure~\ref{fig:approach_access} shows the approachability and accessibility scores at each location on the route (panel A), as well as the spatial distribution of accessibility values at these locations (panel B), assuming $H^* = 2$ $m$, $v = 12$ $ms^{-1}$, and $\zeta = 12$$h$. Panel A shows clear differences between approachability and accessibility scores, which appear to increase as the vessel travels larger distances. At the offshore destination site (36.9° N, 75.35° W), approachability, which does not consider the persistence of a weather window, clearly under-estimates mission feasibility by at least $5$\%. 
Panel B more clearly highlights how accessibility decreases as the vessel travels from the port to the offshore destination, with farther locations corresponding to less frequent persistence of favorable conditions. A risk-averse planner might suggest using $0.85$ as the accessibility score as a representative of mission success; following 
the common adage, ``\textit{a chain is only as strong as its weakest link,}'' suggesting that if any location is not accessible, then the route is not accessible. However, the operational reality is that we do not require $\zeta$-hour persistence at any one specific location in a $\zeta$-hour service route. 
As such, using the offshore site as an indicator of accessibility for a full mission can be overly conservative, resulting in missed opportunities due to imprecise met-ocean assessment. Similarly, using the accessibility score at the port, or the average accessibility across all {sites under-estimates} the likelihood of mission success. 

\begin{figure}
    \centering
    \includegraphics[width=0.9\textwidth]{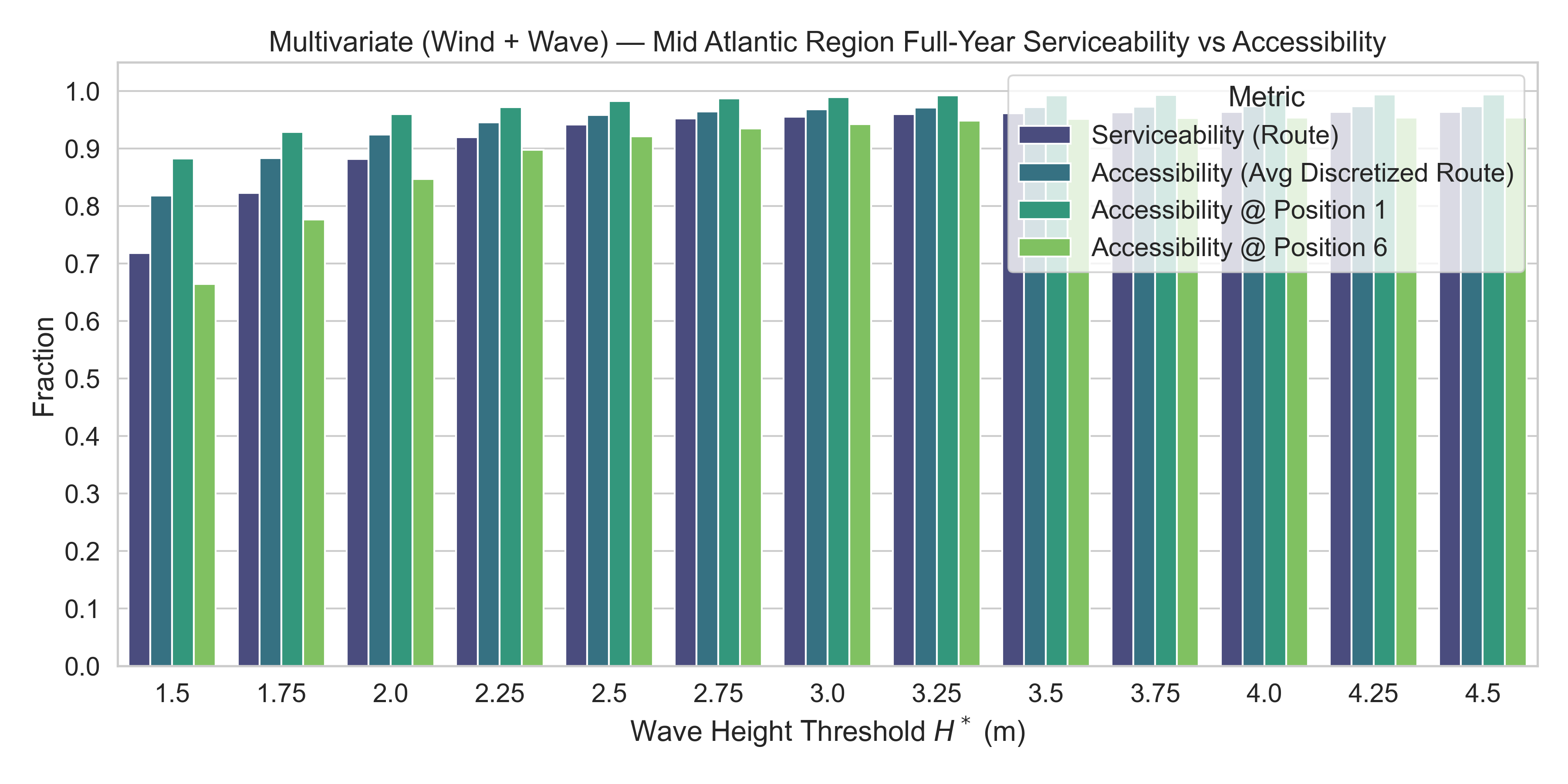}
    \caption{Comparison between route serviceability and various 12-hour accessibility assessments along the Mid-Atlantic route. Position 1 represents the port (36.88, -76.33), while Position 6 is approximately the centroid of the offshore wind farm (36.90, -75.35). Serviceability values are closer to the destination's accessibility, reflecting the extended time spent at the destination, which exhibits harsher met-ocean conditions.}
    \label{fig:service_vs_access_comparison}
\end{figure}

We postulate that serviceability is a logistical metric that mitigates these limitations. Figure \ref{fig:service_vs_access_comparison} shows the route-specific serviceability (computed using the proposed expression in (\ref{eq:serviceability})) across various wave height thresholds, against three other scores that planners can use in absence of a serviceability index: The accessibility at position 1 (port), accessibility at position 6 (destination), and the route-averaged accessibility (averaged over the six positions). 
Notably, the serviceability values fall consistently below the average accessibility of the route, and are closer in magnitude to the destination accessibility than to the port. Although the vessel spends time in both port and transit conditions, a large portion of the mission (three hours) occurs at the offshore site, which tends to have harsher conditions. This proportionally greater exposure to harsh conditions at the destination effectively reduces the overall serviceability metric toward the accessibility levels of the wind farm.

An additional observation is that the accessibility scores for Position 1 increase more rapidly as \(H^*\) increases, reinforcing the idea that calmer and more sheltered port conditions produce consistently high access windows. In contrast, the accessibility at Position 6 curve remains lower at all thresholds, which restricts serviceability, even at higher levels of \(H^*\). As such, we argue that serviceability values are more realistic representations of the vessel's operational journey across the entire mission. 

\subsection{The Impact of Accurate Met-ocean Information on Accessibility and Serviceability}
To ensure precise route-level accessibility and serviceability, access to high-resolution met-ocean information is necessary. This is one of the advantages of our TSR met-ocean model presented in Section \ref{sec:met-ocean_model}. In this Section, we examine the value of this model on accessibility and serviceability assessments. We argue that the use of this met-ocean model is particularly pertinent in the context of serviceability. This is in contrast to solely relying on numerical model data as is typically done in the literature. While numerical model data is extensive in coverage, its spatial and temporal resolution may be insufficient to decipher route-level dynamics, and can exhibit noticeable biases during severe met-ocean conditions, which are most relevant to accessibility and serviceability assessments. 

\begin{figure}
\centering
\includegraphics[width=\textwidth]{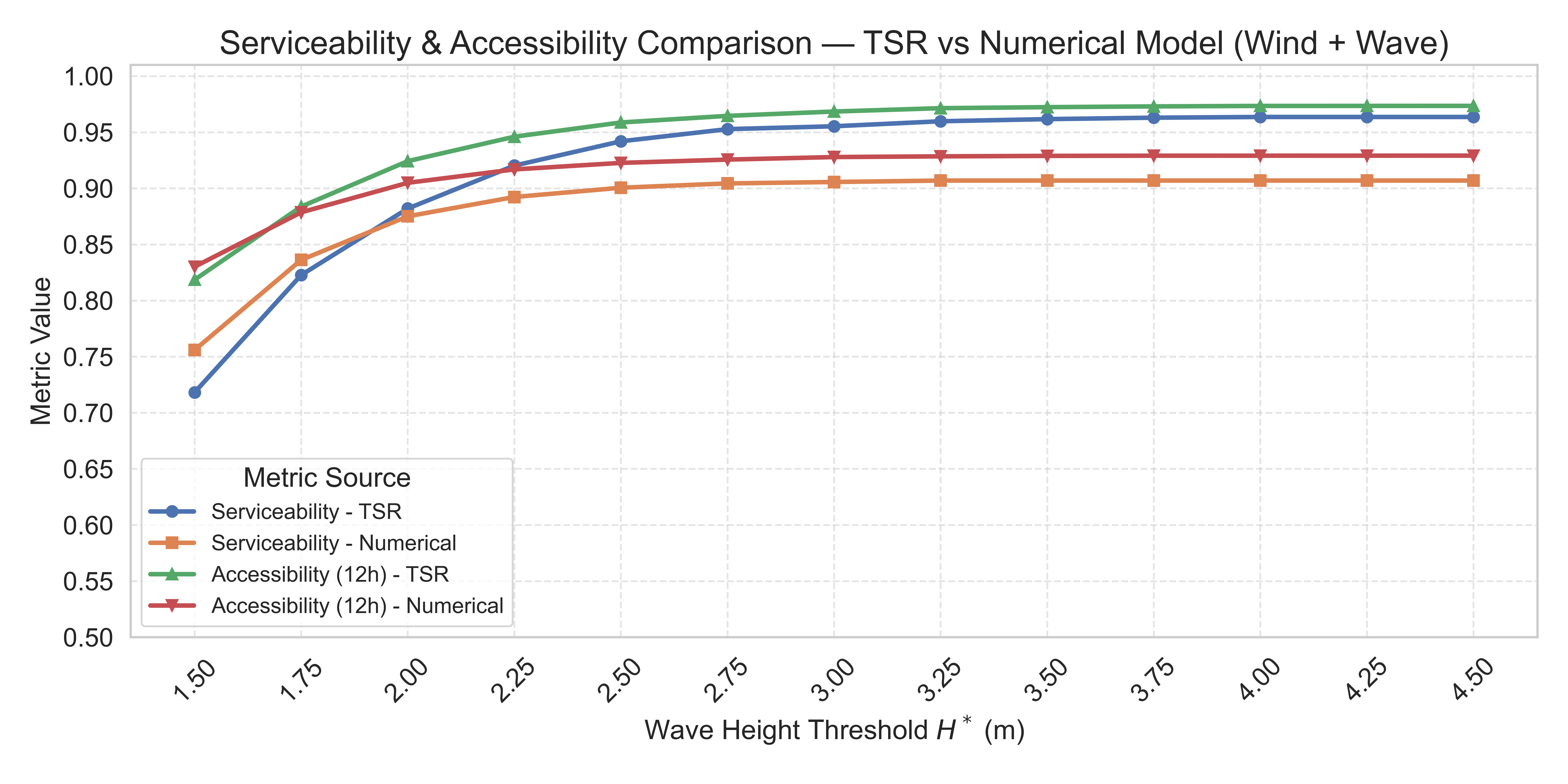}
\caption{Accessibility and serviceability scores derived using the numerical versus the TSR model. Using the numerical model appears to over-estimate accessibility and serviceability at lower thresholds, but under-estimate it at higher thresholds.} 
\label{fig:numerical model_vs_fitted_serviceability}
\end{figure}

To demonstrate this, Figure~\ref{fig:numerical model_vs_fitted_serviceability} compares route-averaged accessibility and serviceability values computed using the numerical model versus that from the fitted TSR model presented in Section 2.1. Looking at the figure, it is clear that at lower thresholds ($H^* \leq 2m$), the accessibility and serviceability values derived from the numerical model are larger in magnitude than those of the TSR model. The difference is more pronounced for serviceability values (about $1$-$4$\%, on average). This is likely attributed to the biases of the numerical model which can over-estimate the likelihood of mission success. 

Interestingly, the relationship between the numerical model and TSR model inverts at higher wave height thresholds ($H^* > 2m$), where the numerical model, on average, yields lower accessibility and serviceability scores than those derived using the TSR. 
We believe that the values derived using the TSR model are more reflective of the underlying true values, considering the improved explanatory power of the model, especially in reducing model error during severe met-ocean conditions, as demonstrated earlier in Section 4.1. As a result, while the numerical model may under-estimate the accessibility and serviceability values around 1.5 to 2 meters, it may, conversely, over-estimate them at thresholds greater than 2 meters. Lower thresholds are more relevant to day-to-day operations conducted by CTVs, whereas higher thresholds are typically invoked for larger vessels that are typically reserved for longer and more complex missions. 

{The differences in serviceability scores observed in Figure \ref{fig:numerical model_vs_fitted_serviceability} raise an interesting question about the impact of such percentage point differences on the economics of offshore energy operations. Definitively answering this question requires a rigorous integration of the serviceability framework into O\&M cost models, which is the focus of subsequent work. Yet, we hypothesize that  embedding high-resolution, route-specific serviceability information into vessel dispatch and O\&M models would likely lead to noticeable economic gains in offshore wind energy operations. This is because such information would enhance the ability of operators to precisely identify opportunistic weather windows of vessel scheduling, thereby lowering logistical requirements and O\&M costs. This hypothesis is supported by the mounting evidence in the literature and practice that higher-resolution met-ocean and logistical information can yield significant gains in offshore operations and overall project availability \cite{petros2021, Petersen2025DataDriven, kolios2023effect}.}
\FloatBarrier


\FloatBarrier
\section{Conclusions and Future Directions}

In this paper, we introduced a methodological framework for accessibility and serviceability assessment, with a regional focus on the U.S. East Coast, to inform logistical planning, development, and O\&M in the offshore wind energy areas therein. The study is motivated by a lack of detailed logistical information that is tailored to the met-ocean behavior in this region, which is a critical input to de-risk any large-scale offshore energy developments. Our work presented a number of unique contributions, including the first publicly available and extensive assessment of logistical accessibility for the U.S. East Coast, as well as the introduction of 
a new logistical metric, \textit{serviceability}, which, unlike common metrics in the literature and practice, accounts for not only the temporal persistence of weather patterns, but also the met-ocean conditions experienced by a vessel along its travel path. We believe that this metric provides a more realistic and operationally relevant depiction of the true likelihood of mission success compared to traditional assessments of approachability and accessibility. Together, these contributions offer a rigorous way to link high-resolution met-ocean information with realistic offshore wind farm construction and O\&M vessel dispatch and scheduling. The methods enable planners to go beyond static, point-based accessibility analysis and instead evaluate entire route feasibility, encompassing important information on both spatial and temporal variability in met-ocean conditions. 

{While the serviceability case study presented in Section 4.2.2. focused on a specific 12-hour operational profile in the Mid-Atlantic region, we note that this is merely illustrative and does not limit the general applicability of the proposed serviceability metric. In fact, the serviceability metric proposed in Section 3.3. is applicable to routes of any duration and complexity, provided that the vessel’s operational profile information is available. This is a reasonable requirement, as these profiles are typically available to mariners and logistical planners for standard offshore operations. This generalizability of the serviceability framework can be leveraged to conduct a comparative assessment of serviceability scores across distinct regional routes with diverse environmental and operational conditions. Such assessment would provide highly valuable information to support detailed project siting decisions, techno-economic analyses, vessel dispatch and O\&M planning, thereby replacing the commonly used spatially-independent metrics which can mis-estimate the true logistical requirements and costs involved in offshore operations. }

Looking ahead, there are multiple possible extensions of this work. For example, future analysis can include more operationally relevant met-ocean variables (e.g., wave period, direction) in assessing accessibility and serviceability. Furthermore, while this paper focused on the Mid-Atlantic region, the framework can be directly generalizable to other areas with significant offshore wind potential, including in the United States (e.g., the Pacific coast and Gulf), as well as elsewhere in the world. Additionally, enabled by the probabilistic nature of the met-ocean model, a possible direction is to develop a data-driven simulation framework which can be directly used to inform logistical planning for offshore wind energy development and O\&M considering environmental and operational uncertainty. {This framework would help establish validated conclusions on how region-to-region and seasonal differences in serviceability scores can impact vessel dispatch decisions, O\&M schedules, and overall project availability.} Another potentially important aspect is to leverage more information in the vessel's operational route to derive important probabilistic insights about route-specific fuel consumption, emissions, and costs. This point is particularly relevant as the interest in vessel electrification continues to ramp up, including for offshore wind applications, but also for other ocean-based industries in the broader Blue Economy. 


\section{Bibliography}
\bibliography{WileyNJD-AMA}%


\end{document}